\documentclass[12pt,preprint]{aastex}

\slugcomment{submitted to the {\it Astrophysical Journal}}
\shorttitle{Dark Matter and NGC 4636}
\shortauthors{Loewenstein and Mushotzky}

\begin{document}

\title{The Nature of Dark Matter in Elliptical Galaxies:{\it Chandra}
Observations of NGC 4636}
\author{Michael Loewenstein\altaffilmark{1}, and Richard F. Mushotzky}
\affil{Laboratory for High Energy Astrophysics, NASA/GSFC, Code 662,
Greenbelt, MD 20771}
\altaffiltext{1}{Also with the University of Maryland Department of
Astronomy}
\email{loew@larmes.gsfc.nasa.gov}

\begin{abstract}
We determine the total enclosed mass profile from 0.7 to 35 kpc in the
elliptical galaxy NGC 4636 based on the hot interstellar medium
temperature profile measured using the {\it Chandra} X-ray
Observatory, and other X-ray and optical data. The total mass
increases as $r^{1.2}$ to a good approximation over this range in
radii, attaining a total of $\sim 1.5\times 10^{12}$ M$_{\odot}$
(corresponding to $M_{\rm tot}/L_V=40$) at 35 kpc. We find that at
least half, and as much as 80\%, of the mass within the optical
half-light radius is non-luminous, implying that NGC 4636 has an
exceptionally low baryon fraction. The large inferred dark matter
concentration and central dark matter density, consistent with the
upper end of the range expected for standard cold dark matter halos,
imply that mechanisms proposed to explain low dark matter densities in
less massive galaxies (e.g., self-interacting dark matter, warm dark
matter, explosive feedback) are not effective in elliptical galaxies
(and presumably, by extension, in galaxy clusters). The composite
(black hole, stars, and dark matter) mass distribution has a generally
steep slope with no core, consistent with gravitational lensing
studies.
\end{abstract}


\section{Introduction}

\subsection{Context}

The presence of extended dark matter halos in late-type galaxies
inferred from analysis of gas and stellar dynamics is
well-established; and, attention is now focussed on comparing the
detailed mass distribution with theoretical predictions of galactic
dark halo structure. In a universe predominantly composed of
non-baryonic matter with specified characteristics, the mass
distribution is determined by the cosmological world model, the
initial density perturbation spectrum, and -- at least on galactic
scales -- the dynamical response to the reconfiguration of the
secondary, baryonic component that occurs during galaxy
formation. Models where non-interacting (except for gravity) cold dark
matter (CDM) constitutes the primary mass component are highly
successful in providing an explanatory framework for a diverse range
of phenomena in the cosmological setting, and are now
standard. However a critical re-examination of the CDM paradigm is
underway, due in large part to its confrontation with measurements of
late-type galaxy mass distributions indicating that dark matter is
less concentrated than expected.

Dark matter density distributions in the centers of collapsed
structures in CDM-dominated model universes increase at least as
steeply as $r^{-1}$ (e.g., Navarro, Frenk, \& White 1997; Ghigna et
al. 2000). However, over a wide range of galaxy luminosity and
morphological type -- including luminous spiral, dwarf, and low
surface brightness galaxies -- observational data favor mass models
with dark matter cores. Some attempt to explain this discrepancy by
proposing an alteration in the nature or perturbation spectrum of dark
matter in such a way that the density distribution evolves to develop
a core, or is initially more diffuse on galaxy scales. Whether any of
these variants offer success over the full range of mass scales from
dwarf galaxies to galaxy clusters, without conflicting with other
astrophysical and particle physics considerations, is an open question
\citep{sil02}. In a separate class of alternative scenarios, the
assumption of the cold, non-interacting nature of dark matter is
retained; but, dynamical effects from other mass constituents during
the galaxy formation epoch reduce the central dark matter
concentration. This feedback may take the form of core heating via
mergers of two systems with supermassive black holes, or gravitational
coupling to powerful outflows in a baryon dominated central
proto-galactic region. \S 5 of the present paper includes details on
the observational situation and these proposed explanations.

The form of the dark matter distribution in elliptical galaxies could
prove decisive in this debate, since these are the most massive
galaxies with the largest supermassive black holes, display strong
evidence for undergoing early outflows and, have high central stellar
densities. Various alternative explanations for the absence of dark
matter cusps in late-type galaxies may very well diverge in their
expectations of whether low central dark matter concentrations occur
in ellipticals as well.

\subsection{Dark Matter in Elliptical Galaxies}

A consensus affirming the presence of dark matter in elliptical
galaxies is finally emerging from a diversity of observational
programs, including optical studies of ionized gas disks \citep{b93},
dynamical modeling of stellar kinematical data (e.g., Gerhard et
al. 2001, Magorrian \& Ballantyne 2001, \S 5.3; but, see Baes \&
Dejonghe 2002), gravitational lensing statistical studies, (e.g.,
Keeton 2001, \S 5.1.2), and analysis of extended X-ray emitting gas
distributions (e.g., Loewenstein \& White 1999). As is the case for
spiral galaxies, dark matter in elliptical galaxies comprises an
increasing fraction of the total mass with distance from the galactic
nucleus, and does not obviously become the dominant constituent until
beyond the optical half-light radius $r_e$.

Progress on constraining the dark matter distribution in elliptical
galaxies was impeded by the absence of fine angular resolution
measurements, over a sufficiently broad dynamical range in radii, of
mass tracers in individual systems.  Data with these qualities are now
obtainable, following the launch of the {\it Chandra} X-ray
Observatory with its capability for imaging and imaging spectroscopy
on sub-arcsecond scales over a $\sim 10'$ field-of-view.  In this
paper we utilize analysis of {\it Chandra} observations of NGC 4636 to
investigate the dark matter distribution in this elliptical
galaxy. Observations and data analysis are described in \S 2, the mass
modeling procedure explained in \S 3, and the resulting dark matter
constraints presented in \S 4. In \S5 we compare these constraints
with theoretical expectations in the context of standard CDM, with
results of gravitational lensing studies, and with observations of
galaxies of other Hubble types -- as well as with expectations for
elliptical galaxies based on various explanations for the low central
dark matter densities in these galaxies. Our conclusions follow in \S
6. A distance of 15 Mpc to NGC 4636 is adopted \citep{t01}.

\section{Observations and Spectral Analysis Results}

NGC 4636 was observed with the {\it Chandra} ACIS-S instrument for 50
ksec on 26 January 2000; an earlier 10 ksec ACIS-I observation is not
discussed here. Spectra are exclusively extracted from the S3 chip in
the four annuli described in Table 1, selected to obtain sufficient
photons for spectral parameter uncertainties to be
systematics-limited. Response and auxiliary response files are derived
using the {\it Chandra} X-ray Observatory Science Center, August 2001,
$-120$ C ACIS S3 chip response products, and {\it Chandra} Interactive
Analysis of Observations software implemented with a script written by
K. Arnaud. Background is subtracted, but is unimportant out to the
radius of the outermost {\it Chandra} spectral extraction annulus
($80''$). Since discrete X-ray binaries are not easily detected
against the bright central diffuse emission inside $\sim 20''$, we did
not explicitly exclude them from the extracted spectra. Instead each
annulus is fit with two-components, including a variable-abundance,
thermal plasma modeled with the APEC code \citep{s01c}, plus a hard
spectral component to account for the binaries. The composite spectra
of X-ray binaries located outside the central $1'$ can be fit by a
$kT=7$ keV bremsstrahlung model (see, also, Angelini et al. 2001 for
NGC 1399); there is no evidence for an associated soft component. We
fix the hard component accordingly; the derived thermal plasma
parameters are insensitive to reasonable variations in the assumed
hard component temperature.

Our derived temperatures, abundances, and column densities are
displayed in Table 1. Also shown are values of reduced-$\chi^2$ for
the best fits; there are no systematic residuals, and most of $\chi^2$
originates at energies $\sim 0.8$ keV.  Temperatures and Fe abundances
are consistent with those derived from the {\it XMM-Newton} Reflection
Grating Spectrometer (RGS) for the central $1'$ (Xu et
al. 2002). Statistical errors in the temperature are 0.01--0.02 keV,
and dominated by systematic errors originating in modeling
assumptions, the plasma code, and the {\it Chandra}
calibration. Additional uncertainties arise from systematic errors in
the S3 response matrix. Prior to the release of the August 2001
response products, these systematic errors precluded reliable detailed
fitting and error estimation. In line-rich sources such as elliptical
galaxies, temperatures derived using the earlier matrices were very
sensitive to the choice of thermal model and unstable to changes in
the assumptions. While new response matrices are considerably
improved, they remain uncertain at energies below 0.7 keV. This
renders the derived column densities and O abundances unreliable, and
explains the inconsistency with the Xu et al. (2002) {\it XMM-Newton}
RGS results for these parameters (for these, and other
calibration-related reasons, we do not quote errors on the abundances
at this time). However, because temperatures derived from CCD spectra
are primarily determined by the ionization balance of Fe and the
effective energy and shape of the Fe-L blended line complex above 0.8
keV, we believe that our temperatures are as reliable as the atomic
physics used to model them. The {\it Chandra} temperatures show good
agreement with those derived from the {\it XMM-Newton} European Photon
Imaging Camera in the regions of angular overlap (J. R. Peterson,
private communication). The APEC code has been tested using RGS
observations of stars and NGC 4636 itself \citep{x02}, and has proved
robust at the level required here. We have not fitted spectra with
multi-temperature models because of the dominance of systematic errors
resulting from calibration and atomic physics parameter uncertainties,
and because the {\it XMM-Newton} RGS fits indicate the presence of a
narrow range of temperatures in the $1'$ RGS beam consistent with a
single-phase gas with the temperature gradient inferred from the {\it
Chandra} data.

\section{Mass Modeling Method and Assumptions}

\subsection{Overview}

The method for constraining the mass profile in NGC 4636 is explained
in Loewenstein \& White (1999). The spherically symmetric equation of
hydrostatic equilibrium is solved for the temperature distribution for
comparison (following emission-weighted averaging) with the observed
profile. The Jeans equation is simultaneously solved, since acceptable
mass models must be consistent with the observed optical velocity
dispersion profile (for some reasonable velocity ellipsoid anisotropy
distribution).  A parameterized gas density distribution is derived
from the {\it Chandra} surface brightness profile. The mass model
consists of a $8\times 10^7$ M$_{\odot}$ central supermassive black
hole \citep{mf01}, a constant mass-to-light component following the
observed optical light profile, and an additional dark matter
component that does not trace the light. The latter is parameterized
by asymptotic density slopes at the origin and at infinity, a scale
length determining the transition from inner to outer slope, and a
normalization.

\subsection{The Stellar Density Distribution}

Analysis of {\it Hubble Space Telescope} ({\it HST}) V- and H-band
images show that the inner stellar surface brightness distribution can
be characterized by a ``Nuker'' model \citep{f97}, while the lower
resolution, but more extensive, ground-based (major-axis) profile can
be fit with a ``Sersic'' model with an exponent that, in this case, is
close to the de Vaucouleurs $R^{1/4}$ law value \citep{ccd}. A
function that provides an excellent fit to the surface brightness over
the entire observed range of projected radii, $R$, is
\begin{equation}
\Sigma(R)=\Sigma_oS(R/r_{\rm break}),
\end{equation}
\begin{equation}
S(x)=2^{(\beta-\gamma)/\alpha}x^{-\gamma}
(1+x^{\alpha})^{(\gamma-\beta)/\alpha}(1+\delta x)^{\beta-3},
\end{equation}
where we adopt the following (V-band) Nuker parameters from
\citet{f97}: break radius $r_{\rm break}=3.21''$, and slopes
$\alpha=1.64$, $\beta=1.33$, and $\gamma=0.13$. The H-band surface
brightness distribution, as measured with {\it HST}/NICMOS confirms
the accuracy of these parameters \citep{qbs,r01}. The surface
brightness at $r_{\rm break}$ is 17.71 V magnitudes arcsec$^{-2}$, and
$\delta^{-1}=175$ produces a smooth transition to the observed Sersic
profile. Equation (1) is numerically deprojected and integrated to
derive, after multiplication by a constant mass-to-light ratio $M_{\rm
stars}/L_V$, the radial stellar mass density and integrated mass
distributions. $M_{\rm stars}/L_V$ generally is a free parameter in
our models and, in effect, subsumes all (dark or luminous) matter
distributed as the optical light.

\subsection{The Gas Density Distribution}

The azimuthally-averaged {\it Chandra} diffuse emission surface
brightness profile in the 0.5--2.0 keV band (see Loewenstein et
al. 2001) is characterized by a flat $\sim 10''$ core, an inflection
in the $\sim 20-40''$ region, and $R^{\sim -2}$ declines both between
the core and the inflection and outside of the inflection. Since the
inflection is associated with non-azimuthal structure in the image
\citep{j01} likely to be out of hydrostatic equilibrium, we do not try
to model the profile in detail but, instead, derive a smooth
characterization of its overall shape. Motivated by energy-balance
arguments \citep{lm87} and the large-scale correspondence of X-ray and
optical surface brightness first noted by \citet{tfc}, we adopt the
following approximation:
\begin{equation}
\rho_{\rm gas}\propto \left(
{{\rho_{\rm stars}/\rho_{\rm stars}(r_{\rm break})}\over 
{1+\rho_{\rm stars}/\rho_{\rm stars}(r_{\rm break})}}\right)^{1/2}.
\end{equation}
Models using a ``double-beta model'' fit yield substantially similar
results, and fits including and excluding the non-azimuthal structure
are consistent at the $\sim 10$\% level.

\subsection{The Dark Matter Density Distribution}

The dark matter distribution of virialized objects in numerical
simulations of structure formation can be characterized by functions
of the ``generalized NFW'' form,
\begin{equation}
\rho_{\rm dark}=\rho_{\rm dark,o}\left({r\over {r_{\rm dark}}}\right)^{-\zeta}
\left(1+{r\over {r_{\rm dark}}}\right)^{\zeta-3},
\end{equation}
where the scale-length $r_{\rm dark}$, for a given cosmological world
model and density perturbation spectrum, is primarily determined by
the virialized mass, but also may vary with formation redshift and
mass accretion history (e.g., Wechsler et al. 2002). Since there is
some theoretical uncertainty in the expected value of $\zeta$, and
since both $\zeta$ and $r_{\rm dark}$ may be altered by coupling of
the halo to an evolving baryonic component, $r_{\rm dark}$ is a freely
varying parameter in our mass models and we consider $\zeta=0$, 1,
3/2, and 2. Integrating equation (4) yields the dark matter mass
distribution,
\begin{equation}
M_{\rm dark}=4\pi\rho_{\rm dark,o}{r_{\rm dark}}^3f(r/r_{\rm dark}), 
\end{equation}
where
\begin{equation}
f(x)=\left\{\begin{array}{ll}
log(1+x)+{2\over {1+x}}-{1\over {2(1+x)^2}}-{3\over 2} & \mbox{$\zeta=0$} \\
log(1+x)-{x\over{1+x}} & \mbox{$\zeta=1$} \\
{2\over 3}log\left(1+x^{3/2}\right) & \mbox{$\zeta=3/2$} \\
log(1+x) & \mbox{$\zeta=2$} 
\end{array}\right.
\end{equation}
In addition to $r_{\rm dark}$ and $\zeta$, the dark matter
distribution is determined by $(M_{\rm dark}/M_{\rm
stars})\left.\right|_{{200r_{\rm break}}}$, the dark-to-luminous mass
ratio at an (optical and X-ray) observationally relevant scale we
choose to be $200r_{\rm break}$ ($10.7'$ or 46.7 kpc) -- a radius
encompassing roughly 90\% of the optical light. We also consider
pseudo-isothermal models with
\begin{equation}
\rho_{\rm dark}=\rho_{\rm dark,o}\left(1+{r\over {r_{\rm dark}}}\right)^{-2},
\end{equation}
and
\begin{equation}
f(x)=x-tan^{-1}x.
\end{equation}

\section{Results}

\subsection{Model Evaluation}

We construct mass models with observed emission-averaged temperature
profiles in projection that match that observed for NGC 4636. In
addition to the {\it Chandra} temperature measurements in four annuli
within $80''$ of the galaxy center (Table 1), successful models are
required to reproduce the observed nearly isothermal ($kT\approx 0.74$
keV) profile in the $1.5-8'$ region. This outer temperature profile
was derived from analysis of {\it XMM-Newton} CCD spectral analysis
(Peterson et al., in preparation), and confirmed by us from
re-analysis of {\it Advanced Satellite for Cosmology and Astrophysics}
({\it ASCA}) data with the APEC plasma model used to derive the {\it
Chandra} temperatures. The statistical errors in the temperatures are
essentially negligible and clearly much lower than systematic
uncertainties, such as those resulting from remaining inaccuracies in
the plasma code and the assumption of spherical symmetry. In this work
we discuss mass models that most nearly reproduce the best-fit
temperature profile (Figures 1, 2, 8), without consideration of
errors, emphasizing the most robust properties of the models. Thus, we
make no strong claims as to the exact functional form of the dark
matter density distribution, but are secure in our conclusions about
the {\it total} mass profile: its statistical uncertainty directly
reflects those in the temperature profile that are on the few percent
level.

\subsection{Constant Mass-to-Light Models}

It is readily apparent that the {\it Chandra} temperature profile
cannot be reproduced with constant mass-to-light ratio models. Figure
1 shows the predictions of models with $M/L_V=8$ and $M/L_V=13.5$ (in
solar units). We confirm that the former value produces the observed
line-of-sight velocity dispersion of $\sim 190-200$ km s$^{-1}$ inside
$r_e/3$ (e.g., Bender, Saglia, \& Gerhard 1994) assuming isotropic
orbits. This failure of constant $M/L_V$ models indicates that dark
matter {\it must} become significant beyond a few kpc.

\subsection{Single Dark Matter Component Models}

For a given central dark matter slope $\zeta$, we find the combination
of $(M_{\rm dark}/M_{\rm stars})\left.\right|_{{200r_{\rm break}}}$,
$r_{\rm dark}$, and $M_{\rm stars}/L_V$ that provides the best match
to the observed temperature profile (for the best-fit model parameters
and characteristics of interest, see Tables 2 and 3,
respectively). Matches for $\zeta=1$ and 1.5 are of comparable
quality, with the latter implying a lower stellar mass-to-light ratio:
$M_{\rm stars}/L_V=3.2$ compared to $M_{\rm stars}/L_V=5.5$. The
best-fit $\zeta=0$ model (with $M_{\rm stars}/L_V=6.6$) is a
marginally worse match. As the assumed value of $\zeta$ is increased,
lower values of $r_{\rm dark}$ and higher values of $M_{\rm
stars}/L_V$ are indicated.  However, as is the case for the constant
$M/L_V$ models, models with $\zeta=2$ are too centrally concentrated
to reproduce the observed temperature profile. The predicted
emission-averaged, projected temperature distributions for the
best-fit $\zeta=1$ and $\zeta=0$ models are shown in Figure 2.

Also shown in Figure 2 is the best-fit $\zeta=0$ model with stellar
mass-to-light ratio fixed at $M_{\rm stars}/L_V=8$ \citep{k00}. This
model is not as successful at reproducing the temperature profile as
models with freely varying values of $M_{\rm stars}/L_V$ (the mean
square temperature deviation is about three times as high and the
innermost temperature badly discrepant; Table 2, Figure 2), a
consequence of the observed combination of relatively steep
temperature rise followed by sharp turnover (naturally, models with
$M_{\rm stars}/L_V=8$ and $\zeta>0$ fare even worse).

While the best-fit stellar mass-to-light profile depends on the
assumed form of the dark matter distribution, the {\it total}
mass-to-light ratio is determined by the temperature profile itself
and is robust, as demonstrated in Figure 3 for the best-fit $\zeta=1,
1.5$, and 0 models. Compared on sufficiently large ($\gtrsim 0.5'$)
scales, the inferred masses are in good agreement with previous
results \citep{mus94,bm97,m98}.

Figures 4a, 4b, and 4c show the dark and luminous mass fraction
distributions for these same models, over the radial range covered by
the {\it Chandra} and {\it XMM-Newton}/{\it ASCA} temperature
profiles. These illustrate that even the high angular resolution {\it
Chandra} data cannot eliminate ambiguity in the relative contributions
of luminous and dark matter inside the half-light radius, $r_e\sim 8$
kpc (e.g., Kronawitter et al. 2000), and in the radius at which dark
matter starts to dominate. However, interesting limits can be derived
-- even for the $M_{\rm stars}/L_V=8$ models that are (at best)
marginally acceptable, $\sim 40$\% of the mass within $r_e$ is dark
and the radius beyond which more than half of the enclosed mass is
dark is $\sim 11$ kpc. These are certainly very conservative lower and
upper limits, respectively. For the best-fit model with a dark matter
core ($\zeta=0$), the dark and luminous component mass fractions are
approximately equal within $r_e$.

Figures 5a, 5b, and 5c show the mass decompositions for the best-fit,
single-component $\zeta=1$, $\zeta=1.5$, and $\zeta=0$ models,
respectively, and further illustrate this tradeoff. Although the
total mass distributions are nearly identical, dark matter is
significant throughout the galaxy for $\zeta=1.5$, but negligible out
to a significant fraction of $r_e$ for the other two slopes.

The average enclosed dark matter densities are shown in Figures 6a and
6b for the best-fit $\zeta=1, 1.5$, and 0 models as well as the
best-fit $\zeta=0$ model with $M/L_V$ fixed at 8. Again, the tradeoff
between allowed dark and luminous matter components is apparent given
sufficient flexibility in the assumed distribution of the former. The
implications of these constraints for the nature of the dark matter in
NGC 4636 is discussed in \S 5.

\subsection{Double Dark Matter Component and Pseudo-Isothermal Models}

We also consider two-component dark matter models composed of
distributions where one component has steep inner density slope
($\zeta_1=1$, 1.5, 2), and the other a flat inner slope ($\zeta_2=0$).
Secondary scale length and normalizations become additional parameters
to be fit for. Best-fit parameters and model characteristics are shown
in Tables 2 and 3: naturally, the additional parameters allow for even
better formal fits (see Figure 2). As before, models with a $\zeta=2$
component are ruled out (the steep-component normalizations are driven
to zero in the fitting process). The range of characteristics (central
dark matter density, dark/luminous mass decomposition, etc.) of such
models is narrower than for the one-component models, as the
additional flexibility afforded by the extra component results in {\it
tighter model-to-model convergence to more exact matches of the
observed temperatures}. The formally best-fit models are composed of
relatively minor flat components with small ($<<r_e$) scale-lengths
and dominant steep components with much larger ($>5r_e$)
scale-lengths. However, fits of comparable quality can be obtained in
models where both dark matter components have scale lengths of the
order of a few $r_e$ or smaller, and where the flat component becomes
significant -- or even dominant -- within $200r_{\rm break}$. If
$M_{\rm stars}/L_V$ is fixed at 8, the normalization of the steep
component is driven to 0 and the models become indistinguishable from
those discussed in \S 4.3 above. The mass-to-light, dark and luminous
mass fraction, and dark matter density distributions are shown in
Figures 3, 4a-c, and 6a-b, respectively, for the best-fit
two-component model with $(\zeta_1,\zeta_2)=(1,0)$ (the curves for the
corresponding model with $(\zeta_1,\zeta_2)=(1.5,0)$ differ very
little from those for the single-component $\zeta=1.5$ model). Figures
7a and 7b show the mass decompositions for these best-fit
two-component models.

The dark matter density profiles discussed above are of the form given
by equation (4), declining as $r^{-3}$ at large $r$ as determined from
numerical simulations. Since reproducing the observed temperature
profile, in particular the steep rise to $\sim 0.75$ keV and sharp
turnover at 10 kpc, requires dark matter to become prominent inside
$r\sim 10$ kpc, the total mass begins to decrease more slowly than $r$
at a radius of this order. As a result, the equilibrium temperature
displays a radial decline initiating just beyond the radius of the
final observed temperature -- a fine-tuned solution that may conflict
with evidence that the gas remains isothermal to much larger radii
\citep{t94,mus94,m98,mat01}. For this reason we also consider
pseudo-isothermal models (equation 5) both as a single component and
in combination with a steeper central density profile dark matter
component. The best-fit models have flatter temperature profiles at
large radii (Figure 8), but similar properties to the $\zeta=0$ models
inside of the observed region under consideration here.

\subsection{Summary of Mass Models, Robust Features}

To summarize, we have considered mass models (in order of increasing
number of parameters) (1) with one dark matter component and $M_{\rm
stars}/L_V=8$, (2) with one dark matter component and freely varying
$M_{\rm stars}/L_V$, (3) with two dark matter components and $M_{\rm
stars}/L_V=8$, and (4) with two dark matter components and freely
varying $M_{\rm stars}/L_V$. The dark matter models have inner density
slopes of 0, 1, 1.5, or 2 and an outer slope of 3; or an inner slope
of 0 and an outer slope of 2 (``pseudo-isothermal''). For models with
two dark matter components, one is assumed to have inner slope 0 and
the other $>0$. Of these models, only those with $M_{\rm stars}/L_V$
fixed at 8 and (even more strongly) those where the dark matter
component has inner slope 2 are clearly ruled out -- and ruled out by
the {\it Chandra} temperature profile alone. The two-component models
provide formally better fits to the observed temperature profile; we
consider this an indication that a more complex dark matter
distribution best explains the data rather than as evidence of a true
two-component dark matter configuration (although this is
possible). The derived integrated mass distribution on $>100$ pc
scales is {\it nearly model-independent} (more so for the
two-component models), but the dark/luminous matter breakdown is less
robust.

$M_{\rm stars}/L_V$ ranges from 2.4--4.0 (2.4--6.6) for acceptable
best-fit two-component (all acceptable best-fit) models. $M_{\rm
tot}/L_V$ is much more narrowly proscribed, ranging from 6.5--6.7
(6.5--7.0), 14.0--14.3 (13.6--14.4), and 45.6--47.0 (44.8--47.5), at
$r_e/6$, $r_e$, and $6r_e$, respectively (using $r_e=r_{\rm
break}/0.03=7.8$ kpc). The dark matter fraction is $>0.7$ within $r_e$
and $>0.9$ within $6r_e$ for acceptable best-fit two-component
models. The similarity of the radius where dark matter starts to
dominate and the stellar scale-length is the elliptical galaxy
counterpart to the ``disk-halo conspiracy'' and may be due to feedback
on the dark matter distribution from the luminous matter as it evolves
on a dynamical timescale \citep{lcs}.

The average dark matter density within $r_{\rm break}$ (233 pc) is 5.4
M$_{\odot}$ pc$^{-3}$ for the best-fit two-component model with inner
dark matter exponent $\zeta=1.0$, and 16 M$_{\odot}$ pc$^{-3}$ for
$\zeta=1.5$, 0.18 M$_{\odot}$ pc$^{-3}$ for the best-fit
single-component models with cores, and 0.06--0.09 M$_{\odot}$
pc$^{-3}$ for the best-fit cored single-component models with $M_{\rm
stars}/L_V$ fixed at 8.

\section{Discussion}

We compare our constraints on the dark matter distribution in NGC 4636
with theoretical expectations, results from gravitational lensing
studies, and observational results for other types of galaxies.

\subsection{The Dark Matter Distribution in NGC 4636}

\subsubsection{Comparison with CDM Simulations}

As our emphasis is on the luminous region of NGC 4636, we do not
constrain the dark matter density slope at very large radii, although
we present successful models with dark matter density declining both
as $r^{-2}$ and $r^{-3}$ for large $r$. Note that the scale lengths
for many of the successful models are comparable to the outermost
region that we consider and the dark matter density slopes are, on
average, shallower than $r^{-2}$ on the galaxy scale.

{\it Our derived constraints on the dark matter distribution are
consistent with the unmodified predictions of CDM numerical
simulations.} For the purpose of comparison, let us consider our
best-fit two-component models (qualitatively, the discussion does not
depend on this). The best-fit halos have virial radii and masses (for
an overdensity, with respect to the critical density assuming $H_o=70$
km s$^{-1}$ Mpc$^{-1}$, of $\sim 100$ as appropriate for a
$\Lambda=0.7$, $\Omega=0.3$ cosmology) of $\sim 650$ kpc and
$1.5\times 10^{13}$ M$_{\odot}$, respectively (these values are robust
provided the dark matter density declines as $r^{-3}$ at large $r$;
larger values are implied if the slope is flatter). The asymptotic
stellar visual luminosity is $7.7\times 10^{10}$ L$_{V\odot}$,
corresponding to a total stellar mass of $\sim 2-6\times 10^{11}$
M$_{\odot}$, with values at the lower end of this range favored by the
stellar mass-to-light ratios in our best-fit models. In addition, the
total mass in hot gas out to $\sim 300$ kpc is $\sim 5\times 10^{11}$
M$_{\odot}$ (Matsushita et al. 1998, adjusting for our different
assumed distance). Thus, the baryon fraction is $<0.07$ -- about half
the value derived from Big Bang cosmology for $H_o=70$ km s$^{-1}$ and
$\Omega=0.3$. This is not surprising -- elliptical galaxies are
expected to lose much of their initial baryonic content in
proto-galactic outflows (e.g., Loewenstein \& Mushotzky 1996). The
virial temperature corresponding to this dark matter virial mass, 0.3
keV, is well below that measured in NGC 4636.

The dark halo concentration can be characterized by a number of
non-parametric quantities. One such measure is $c_{1/5}$, the ratio of
the virial radius to the radius enclosing one-fifth the virial
mass. We derive values $\sim 9.5$, more concentrated than the average
predicted by numerical simulations, but within the considerable
scatter theoretically predicted at this mass scale (Avila-Reese et
al. 1999, Wechsler et al. 2002).

The non-parametric central dark matter density parameter
$\Delta_{V/2}$ defined by Alam, Bullock, \& Weinberg (2002) -- the
density, in units of the critical density, evaluated at the radius
$r_{V/2}$ where the dark matter circular velocity drops to half of its
maximum value (400--450 km s$^{-1}$ in NGC 4636) -- can be used to
compare the central regions of halos of different mass with each other
and with theoretical expectations. The values for our models are
displayed in Table 3. For our models with freely varying mass-to-light
ratio, $\Delta_{V/2}\sim 3\times 10^6$ -- greater than the average
predicted in standard $\Lambda$-CDM scenarios \citep{abw} or estimated
in less luminous, dark matter dominated galaxies (hereafter
DMDGs). Likewise, $r_{V/2}$ ($\sim 2$ kpc) is correspondingly smaller.

By comparing to dark-matter-only simulations, we ignore the effects of
interaction between the baryonic and non-baryonic components. The
evolution of the dissipational, baryonic component may result in
contraction of the dark matter caused by infall-induced compression
\citep{r91,kw01} or, alternatively, to expansion via outflow (Binney,
Gerhard, \& Silk 2001; but, see Gnedin \& Zhao 2002) or energy
transfer via dynamical friction \citep{esh}. In fact both kinds of
processes are expected, and could effectively cancel each other out
(Lia et al. 2000).  The evolution of binary black holes following a
merger could also reduce the central dark matter density \citep{m02};
but only on a scale corresponding to the stellar break radius ($\sim
230$ pc for NGC 4636), and would need to be followed by an additional
mechanism (starburst-induced outflow following the merger?)  to effect
the dark matter on scales relevant to our measurements. Further
evaluation of the NGC 4636 mass profile in this context awaits more
comprehensive theoretical treatment.

\subsubsection{Connection with Gravitational Lensing Surveys}

There are a number of statistical studies of the observational effects
of gravitational lensing by elliptical galaxies. For simplicity, most
of these (e.g., Kochanek 1994, McKay et al. 2002) approximate the
total galaxy potential as arising from a singular isothermal sphere
(SIS), and steep ($\sim r^{-\sim2}$) mass profiles are indicated by the
distribution of image separations \citep{a00,pm00,kw01,c01,lo02} and
the absence of detectable odd images \citep{rm01}, and consistent with
galaxy-galaxy lensing statistics \citep{bbs,s01a,wil01} and
gravitational time delay measurements (e.g., Kochanek 2002). In NGC
4636, the dark matter density profile is shallower than $r^{-2}$
inside 10 kpc. Figure 9 compares the best-fit two-component --
$(\zeta_1,\zeta_2)=(1,0)$ -- model with an SIS of identical virial
mass ($\sigma_{\rm SIS}=224$ km s$^{-1}$). The total mass
distribution, while poorly matching the stellar distribution in shape,
does not strongly deviate from a comparable isothermal sphere for
$r\gtrsim 1$ kpc (in the region at which it is best-constrained,
$M\propto r^{\sim 1.2}$), but flattens from 1 kpc into 100 pc where
the nuclear black hole starts to dominate the total mass. \citet{cy99}
assume that the mass profile in elliptical galaxies turns over at
about this radius and find good agreement with the numbers and
separations of lenses for a low-density universe with a cosmological
constant $\Lambda=0.7$. Since the stellar density profiles in galaxies
less luminous than $L_*$ that dominate the lensing signal in
statistical studies tend to be considerably steeper than for a galaxy
as luminous as NGC 4636 \citep{f97}, an SIS model may be an adequate
representation of the mean mass distribution for some purposes;
however, more realistic composite models are clearly preferable --
e.g., $M\propto r^{\sim 1.2}$ distributions would imply $\sim 20$\%
higher values of the Hubble constant for a given gravitational time
delay measurement \citep{ko02}. Note that, for NGC 4636, the ratio of
equivalent SIS velocity dispersion to central stellar velocity
dispersion is greater than derived assuming a pure isothermal sphere,
though less than the original assumed value of $1.5^{1/2}$
\citep{k94}.

\cite{k01} does, in fact, consider composite (star$+$halo) mass models
with adiabatic contraction (see, also, Porciani \& Madau 2000;
Kochanek \& White 2001; Jimenez, Verde, \& Oh 2002; Seljak 2002) and,
from strong gravitational lens statistics, infers a dark matter
fraction within $r_e$ of about one-third for the average elliptical
galaxy, compared to 0.7--0.8 in our best-fit two-component models, 0.5
for the best-fit one-component model with dark matter core, and 0.4
for the less-well-fitting fixed mass-to-light ($M_{\rm stars}/L_V=8$)
model. Evidently, NGC 4636 is an unusually dark matter dominated
galaxy (see below; Wilson et al. 2001), which implies a {\it cosmic
variance in the baryonic/non-baryonic ratio} (at least within the
optically visible part of the galaxy; see, also, Kronawitter et
al. 2000) that may be either fundamental, or a result of variations in
the magnitude of baryon-driven reshaping of dark matter halos or of
baryonic mass loss.

\subsection{Implications for Self-Interacting Dark Matter and Other 
Alternative Models}

As discussed above, the dark matter distribution in NGC 4636 is not as
concentrated as the stars or as a singular isothermal sphere, but is
consistent with the results of CDM halo formation simulations,
unaltered by subsequent coupling to the contraction or expansion of
the baryonic component.  This seems not to be the case for many other
types of galaxy.

Possible conflicts between CDM predictions and observations of central
dark matter densities were first pointed out by \citet{m94},
\citet{fp94}, and \citet{b95} for dwarf spiral galaxies. More recent
work revealed further evidence of this discrepancy based on rotation
curves of low-surface brightness and dwarf irregular galaxies
\citep{m99,ccf,db01,ma02}. Claims that inner rotation curves or the
zero-point of the Tully-Fisher relation imply that dark matter is less
concentrated than predicted were made for spiral galaxies
\citep{bac,bs01}, including the Milky Way \citep{ns00a,be01}. It is
notable, however, that contradictory conclusions were reached, even
for low-surface brightness \citep{vdb00,jvo} and dwarf \citep{vs01}
galaxies that are dark matter dominated at all radii, as well as for
the Milky Way \citep{kzs} and other disk galaxies \citep{jvo}.

Estimates of the central dark matter density in DMDGs (Kravtsov et
al. 1998, C\^ot\'e et al. 2000, Firmani et al. 2001b, de Blok et
al. 2001, Marchesini et al. 2002), as well as spiral (Blais-Ouellette
et al. 2001, Borriello, \& Salucci 2001; but, see Klypin et al. 2002)
galaxies cluster around 0.02 M$_{\odot}$ pc$^{-3}$, albeit with a
large scatter. These are based on fitting models with flat dark matter
density cores; fitted core radii are $\sim 1-10$ kpc. Similar central
densities were inferred for clusters of galaxies \citep{wx00,abg};
although, it is less clear whether dark matter distributions that are
less concentrated than predicted by CDM are implied (Arabadjis et
al. 2002), since dark matter is expected to be more diffusely
distributed on cluster, than on galaxy, mass scales. Moreover, density
cusps in clusters may be required to explain their lensing properties
(Smith et al. 2001b, Molikawa \& Hattori 2001; Wyithe, Turner, \&
Spergel 2001).

A giant elliptical galaxy (virial mass $>10^{13}$ M$_{\odot}$), NGC
4636 can bridge the gap in mass scale between previously studied
galaxies ($<10^{11}$ M$_{\odot}$ for DMDGs, $\sim 10^{12}$ M$_{\odot}$
for spirals) and clusters ($>10^{14}$ M$_{\odot})$. Since unlike
clusters and DMDGs, ellipticals are dominated by baryons in their
nuclei, the inferred dark matter density is model dependent. For the
best-fit two-component models presented above, the average dark matter
density (that is roughly equal to the local density) within the
innermost measured {\it Chandra} temperature (corresponding to the
effective resolution of our mass determination) is $\sim 2$
M$_{\odot}$ pc$^{-3}$ (Figure 6a) -- about two orders of magnitude
larger than is typical in the other types of galaxies discussed
above. The maximum values in our acceptable models extend a further
factor of two higher.  For the best-fit models with a dark matter core
the central density is about an order of magnitude lower at $\sim
0.16$ M$_{\odot}$ pc$^{-3}$ (representing $\sim 2$\% of the total
density); the corresponding value if $M_{\rm stars}/L_V=8$ is reduced
by a further factor of two.  This conflict between the central dark
matter density in NGC 4636 and the surprisingly low values previously
found for other galaxy types is robust, provided we maintain the
reasonable assumptions that the dark matter density profile
monotonically decreases outward, and the stellar mass-to-light ratio
does not steeply decline with radius.

\subsubsection{Self-Interacting Dark Matter}

If cold dark matter is self-interacting (SIDM), an initially cuspy
density distribution with its positive velocity dispersion gradient
may develop a core. \citet{ss00}, and others, proposed SIDM as a
possible explanation for previously measured low central dark matter
density cores. For a core to form in a Hubble time in a DMDG, the
specific cross section, $\sigma_{\rm dm}$, must be sufficiently high
for several collision times to elapse (e.g., Dav\'e et al. 2001). This
lower limit is $\sim 10^{-24}-10^{-23}$ cm$^2$ GeV$^{-1}$ for the
typical density of 0.02 M$_{\odot}$ pc$^{-3}$. If the cross section
$\gtrsim 10^{-22}$ cm$^2$ GeV$^{-1}$, energy transfer to the outer
regions will result in core collapse for an isolated halo \citep{m00},
although dynamical heating from continual infall may forestall this.

The nature of the dark matter particle interactions may be such that
either $\sigma_{\rm dm}\propto 1/v_{\rm dm}$ or $\sigma_{\rm dm}$
constant with $v_{\rm dm}$ is possible at the energies of interest,
where $v_{\rm dm}$ is the mean dark matter particle velocity (Hui
2001, Hennawi \& Ostriker 2002). If clusters of galaxies also have
dark matter cores of comparable central density (Wu \& Xue 2000,
Arabadjis et al. 2002), a $1/v_{\rm dm}$ dependence is required for a
single SIDM particle to account for both (D'Onghia, Firmani, \&
Chincarini 2002; Yoshida et al. 2001; Miralda-Escud\'e 2002). For
$\sigma_{\rm dm}=\sigma_{\rm dm}(v_o)({v_o}/{v_{\rm dm}})$, with
$v_o=100$ km s$^{-1}$, $\sigma_{\rm dm}(v_o)\sim 10^{-24}$ cm$^2$
GeV$^{-1}$ is required \citep{dfc,d01}, although Arabadjis et
al. (2002) derive an upper limit that is an order of magnitude lower.

Isothermal cores scale such that $\rho_{\rm iso}r_{iso}\propto
\sigma_{iso}^2$, where $\rho_{\rm iso}$, $r_{iso}$, and $\sigma_{iso}$
($\propto v_{\rm dm}$) are the central density, core radius and
velocity dispersion, respectively. Thus, clusters are expected to have
$\sim 10\times$ larger cores than DMDGs, consistent with fitting cored
models to X-ray observations (Wu \& Xue 2000, Arabadjis et
al. 2002). Since our fits for models with cores have $\sim 10\times$
the central density of DMDGs at a velocity scale several times higher,
core radii are expected to be of the same order. We, indeed, find this
to be the case: the core radius is 4.5 kpc for the best-fit
pseudo-isothermal model, 12 kpc for the cored model with asymptotic
$r^{-3}$ slope.  However the higher density and velocity imply that
the SIDM particles in ellipticals undergo an order of magnitude (or
more if $\sigma_{\rm dm}$ is constant with $v_{\rm dm}$) greater
number of collisions than in DMDGs of comparable age. Whether this
would induce core collapse must be addressed by realistic galaxy
formation simulations that account for energy input from mergers and
infall (note that the dark matter in NGC 4636 does not have the
$r^{-2}$ form characteristic of cores that undergo dynamical collapse;
Burkert 2000, Kochanek \& White 2000).

Simulations with SIDM performed to date \citep{d01,dfc} are consistent
with the scaling proposed by \citet{me02} that essentially assumes a
narrow range in the total number (on the order of a few) of
collisions.  For, $\sigma_{\rm dm}\propto v_{\rm dm}^{-\kappa}$, this
implies
\begin{equation}
r_{\rm iso}\propto \sigma_{\rm iso}^{{3-\kappa}\over 2}\tau^{-{1\over 2}},
\end{equation}
and
\begin{equation}
\rho_{\rm iso}\propto \sigma_{\rm iso}^{\kappa-1}\tau^{-1},
\end{equation}
where $\tau$ is the age of the dark matter halo. Thus (assuming
$\kappa\le 1$), $\rho_{\rm iso}$ is expected to be of the same order
or smaller for elliptical galaxies compared to DMDGs and spirals,
{\it contrary to our models with dark matter cores}.

This leads us to consider interpreting our best-fit models where the
inner dark matter density is cuspy, in the context of SIDM models, as
dark matter cores that subsequently develop cusps as a result of
adiabatic contraction during the collapse of the baryonic component.
The central phase space density $Q\equiv\rho(2\pi\sigma^2)^{-3/2}$ in
NGC 4636 for these models, derived from solving the Jeans equation for
the dark matter, is $\sim 10^{-8}$ M$_{\odot}$ pc$^{-3}$
(km/sec)$^{-3}$ (and rising) at the innermost (X-ray) observed radius
of 700 pc. This is well above the SIDM predictions found by
\citet{d01} to be consistent with the $Q\propto v_{\rm dm}^{-3}$
scaling expected from a ``quiet'' non-dissipational merging hierarchy
(Dalcanton \& Hogan 2001; see, also, Madsen 2001). Since $Q$ is
conserved in adiabatic contraction, this explanation fails.

To summarize, the central density in models of the mass distribution
in NGC 4636 with dark matter cores are higher than expected if all
galaxies develop cores due to dark matter self-interaction.  If the
SIDM cross section, and its velocity dependence, are such that it
reduces the dark matter concentration only on mass scales well below
that of NGC 4636 then galaxy clusters are also unaffected.  Moreover,
this is inconsistent with expectations of the galaxy formation merging
hierarchy unless these cores form rather late.  Therefore, {\it the
NGC 4636 data do not support SIDM models for any cross section
velocity dependence.}

\subsubsection{Warm Dark Matter}

The menagerie of other proposed alternative dark matter candidates
includes warm dark matter \citep{hd00,cav,ens,bot,a01,dh01,abw},
self-interacting scalar fields \citep{p00,rt00}, annihilating dark
matter \citep{kkt}, decaying dark matter \citep{cen01}, fuzzy cold
dark matter \citep{hbg}, repulsive dark matter \citep{goo00}, and
mirror matter \citep{mnt}. Our limit on the central dark matter
density in NGC 4636, and conclusion that if the dark matter mass
distribution does have a core it is of the same order as in the much
less massive DMDGs rule out a number of models, e.g.  those that
predict a universal central density (e.g., annihilating dark matter as
in Kaplinghat et al. 2000), a self-similar reduction in concentration
(e.g., decaying dark matter as in Cen 2001), or a dark matter core
radius that decreases with mass (annihilating dark matter as in Craig
and Davis 2001, fuzzy cold dark matter as in Hu et al. 2000).

The most fully developed of these other alternatives is warm dark
matter (WDM); however, phase-space constraints must be considered. As
summarized above, the central dark matter phase space density $Q$, for
a merging hierarchy of equilibrium systems, should scale as $v_{\rm
dm}^{-3}$ \citep{dh01}. The warm dark matter particle mass, $m_{\rm
wdm}$ (if fermionic) sets a maximum attainable value $Q=1.6\times
10^{-4}(m_{\rm wdm}/1 {\rm keV})^4$ M$_{\odot}$ pc$^{-3}$
(km/sec)$^{-3}$. Since a mass $m_{\rm wdm}\sim 1$ keV is consistent
with the value of $Q$ inferred for dwarf spheroidal galaxies
\citep{hd00,l02}, these may be identified with the initial seeds in
the merging hierarchy and provide the normalization for the phase
space scaling. Once again, NGC 4636 is in conflict with the expected
scaling.

For our models with dark matter cores, the phase space density is
acceptable but the core density (radius) too high (low): as with SIDM,
constant (or decreasing) central dark matter density with mass is
expected. For our models with dark matter cusps, the central dark
matter phase space density is too high. However, while \citet{dh01}
argue that the phase space scaling holds even if the smallest galaxies
are not themselves part of the merging hierarchy, there may be a
fundamental difference in the cosmogony of galaxies above and below
the WDM filtering mass $\sim 10^9-10^{10}$ M$_{\odot}$ (Avila-Reese et
al. 2001; Bode et al. 2001) that might lead one to expect that the
decrease in concentration compared to CDM models should apply only to
these low mass systems. Thus WDM remains a viable alternative from an
astrophysical perspective.

\subsubsection{The Initial Perturbation Spectrum}

While WDM introduces an effective change in the proto-galactic
fluctuation spectrum, other more fundamental departures from standard
CDM are possible. A tilted cold dark matter spectrum with shape
determined by Lyman $\alpha$ forest data leads to reduced central dark
matter densities on all mass scales \citep{abw}. The predictions of
this model that $\Delta_{v/2}<10^5$ at elliptical galaxy scales is
{\it not} consistent with our constraints on NGC 4636.  Other models
that propose reducing the amount of small-scale power (e.g.,
Kamionkowski \& Liddle 2000) may similarly suffer.

\subsubsection{Explosive Feedback}

Dark matter cusps in dwarf spiral and low surface galaxies can be
erased by coupled expansion to a sufficiently powerful
starburst-driven baryonic outflow \citep{nef,gs99,vdb00,bgs2}. For a
constant energy input per stellar mass, this introduces a natural mass
scale delineating systems where this mechanism reduces the dark matter
concentration from those where it is ineffective in doing so. If the
feedback efficiency is tuned to reproduce the Tully-Fisher relation
for disk galaxies, dwarf galaxies are expected to have dark matter
cores. Cores in disk galaxies are more problematic
\citep{ns00b,vdb00}; while giant elliptical galaxy dark matter
distributions would be relatively unaffected as indicated by the
constraints presented here. These calculations implicitly assume that
the feedback occurs after assembly of the dark halo; the conclusions
may differ if the same feedback operating in dwarf galaxies also
occurs in proto-galactic fragments that eventually merge to form
ellipticals.

\subsection{Stellar Kinematics, the Stellar Mass-to-Light Ratio, and
the Dark Matter Fraction}

Published measurements of the central stellar line-of-sight velocity
dispersion in NGC 4636 mostly range from $\sim 180-230$ km s$^{-1}$
\citep{m95,cmp}, values we find consistent with solutions of the Jeans
equation for the stars that include the mass models discussed
above. The velocity dispersion profile is measured only as far out as
$\sim 30''$ ($<r_e/3$; Bender et al. 1994, Caon et al. 2000) and is
roughly constant, except perhaps for a rise inside a kinematically
decoupled core ($<2''$; Caon et al. 2000). Evidence for dark matter
within $r_e$ based on stellar kinematics was presented by \citet{kr85}
(using data in Davies 1985) and, more recently, by \citet{sbs} and
\citet{k00}. \citet{k00} infer $M_{\rm tot}/L_V=9.9$ (after adjustment
for bandpass and distance) at $34''$, compared to $M_{\rm
tot}/L_V\approx 8.3$ in our best-fit models (where $M_{\rm
stars}/L_V=3-4$).

\citet{g01} restrict their consideration to the properties of minimal
dark halo models, as their central mass-to-light ratios are generally
consistent with those inferred from the stellar population, assuming a
local Galaxy initial mass function as described by \citet{kr01}. Their
mass models, that assume a logarithmic dark matter potential with a
core, more closely resemble the one component $\zeta=0$ models
described above (especially that with $M_{\rm tot}/L_V$ fixed at 8)
than the models that best describe the X-ray data. NGC 4636 may be
exceptionally dark matter dominated inside $\sim r_e$ (or,
alternatively, have an exceptionally high magnetic pressure; Brighenti
\& Mathews 1997); however, a comparison of elliptical galaxy central
dynamical mass-to-light ratios with the recent estimates of
\citet{k02} suggest that dark matter may generally be more significant
at all radii than usually presumed (see, also, Seljak 2002). The value
of $M_{\rm stars}/L_V=5.4$ estimated by \citet{g01} based on the
stellar population is significantly less than the dynamically inferred
total ratio. Since a two-component dark matter model with
$(\zeta_1,\zeta_2)=(1,0)$ and $M_{\rm tot}/L_V$ fixed at 5.4 is only
marginally less acceptable than the best-fit two-component model
(Table 2), the mass distribution is consistent with the stellar
population. NGC 4636 is an outlier from the average relationships
presented in \citet{lw99} -- the ratio of central stellar to global
gas temperatures ($\beta_{\rm spec} \equiv \mu m_p\sigma^2/kT$) of
0.34 is indicative of an exceptionally dominant (within $r_e$) dark
halo (see, also, Brighenti \& Mathews 1997); and, the upper limit on
the baryon fraction (\S 5.1.1) is lower than may be typical
\citep{gs02}. Interestingly, NGC 4636 also lies in the mass loss part
of structural parameter ($\kappa-$) space \citep{bbf}, which also may
indicate a low baryon fraction. A more integrated joint analysis of
the X-ray and optical data that considers dark matter distributions
with and without cores, and relaxes the assumption that stars dominate
the total mass at small radii, could prove illuminating, perhaps
shedding light on the apparently kinematically decoupled core (Caon et
al. 2000) and eliminating the need for an anomalously tangential
velocity dispersion ellipsoid \citep{k00}. More extended optical
spectroscopy is clearly desirable, as well. Finally, we note that if
some of the dark matter is baryonic, then it must be distributed in a
less concentrated manner than the starlight -- unlike what might be
expected for stellar remnants.


\section{Summary and Concluding Remarks}

We have constructed mass models of the elliptical galaxy NGC 4636
based primarily on the density and temperature distributions of the
hot gas measured with the {\it Chandra} X-ray Observatory and the
stellar light density distribution measured in the optical. Secondary,
observational inputs are the {\it XMM-Newton}/{\it ASCA} X-ray
temperature profile and the stellar velocity dispersion profile. We
derive accurate constraints on the total mass distribution from
0.7--35 kpc.  The total mass increases as $r^{1.2}$ to a good
approximation over this range in radii, attaining a total of $\sim
1.5\times 10^{12}$ M$_{\odot}$ (corresponding to $M_{\rm tot}/L_V=40$,
Figure 3) at the outermost point we consider.

Of the models we investigate, the temperature profile is most
accurately fit using a dark matter distribution that consists of
two-components -- one with a flat core and one of the generalized NFW
form with a cusp (equation 4).  We consider this an indication that
the dark matter distribution may be more complex than a single
component following equation (4). There is no unique mass
decomposition into luminous and dark matter components (see Figures 5
and 7); steeper assumed values of the central dark matter density
slope ($\zeta$) imply lower stellar mass-to-light ratios (such a
degeneracy exists even for dark matter dominated dwarf galaxies; van
den Bosch \& Swaters 2001). However, constant mass-to-light models and
models with $\zeta\ge 2$ are too steep, and are ruled out by the {\it
Chandra} data. $M_{\rm stars}/L_V\le 6.6$ is indicated from fitting
models with dark matter cores, and models with $M_{\rm stars}/L_V\le
5.5$ -- consistent with the stellar population in NGC 4636 -- are
favored.  While a wide range of central dark matter densities and
dark-to-luminous mass ratios are allowed (Figure 4 and 6), all of
these are sufficiently restrictive so as to have important
implications. The mass profile significantly departs from that of the
light inside a few kpc ({\it i.e.} $\sim 0.5\times$ the half-light
radius $r_e$). The dark matter fraction is $\sim 0.5-0.8$ within
$r_e$, and therefore on the high concentration end of expectations for
CDM halo formation models. Moreover, the central dark matter density
is at least an order of magnitude -- and possibly more than two orders
of magnitude (at $\sim 2$ M$_{\odot}$ pc$^{-3}$, based on the best-fit
two-component models) -- greater than reported as typical for a
variety of spiral and less luminous, dark matter dominated galaxies.

Non-parametric measures of the dark matter central density also
indicate that dark matter in NGC 4636 is more concentrated than
average for a standard CDM halo, though within the scatter. Thus, the
effects of adiabatic contraction and expansion via explosive feedback
are negligible, effectively cancel each other out, or operate on a
halo that initially differs from those computed for CDM halos.

The dark matter distribution in NGC 4636 is at odds with many
alternative models designed to explain low central densities in
galaxies of other Hubble type. It is instructive to consider two
classes of models from this perspective -- those with flat dark matter
cores, and those with cuspy dark matter cores (that provide better
fits to the X-ray data) interpreted as contracting from initially flat
cores due to baryonic infall. For the flat-core models, the high
central dark matter mass density and large cores are contrary to the
expected scaling relations for self-interacting dark matter and other
models where dark matter structure is driven by dark matter particle
interaction. For the cuspy-core models, the central phase-space
density, conserved during adiabatic contraction, is too high. Models
where the flattening of dark matter cores occurs only at relatively
low mass scales (perhaps, at the low surface brightness galaxy scale;
van den Bosch et al. 2000) -- including some involving explosive
feedback and warm dark matter -- remain most viable. If this
transitional mass is indeed on the giant elliptical galaxy scale (or
below), cuspy dark matter distributions in galaxy clusters are
implied.  If the results for dark matter dominated, low mass galaxies
are correct, there is a reversal from the trend naively expected from
a standard bottom-up hierarchy where less massive objects collapse
earlier in a more dense universe and retain that higher density.

Perhaps some modification to the standard CDM scenario is required. It
should be noted, however, that considerable scatter in dark matter
concentration is predicted for a given mass range -- particularly on
galaxy scales. Moreover, this scatter is not purely random, but
represents the effects of variations in age and assembly history
\citep{w02}. This introduces a bias such that galaxies of the types
where low central dark matter densities are derived may represent
relatively recently formed systems, and/or particularly fragile
galaxies that have experienced relatively tranquil assembly
histories. Conversely, giant elliptical galaxies such as, or perhaps
particularly, NGC 4636 may have the highest formation redshifts and
the most prominent merger histories.

Gravitational lensing probabilities are sensitive to the inner total
mass density slope \citep{wb00,bks,tc02,wts,ots,lo02}, and explainable
if the total mass in clusters follows an NFW profile and that in
elliptical galaxies a singular isothermal sphere profile
\citep{lo02}. This is consistent with our results, once one takes into
account the stellar contribution. If one is to use lensing statistics
to constrain cosmological parameters, the correct composite mass model
-- derived empirically as we do here -- should be used. Moreover,
these studies assume a constant dark matter distribution for galaxies
of a given optical luminosity, while our results indicate the
likelihood of a significant cosmic variance.

In fact, we suggest that NGC 4636 is an exceptionally dark matter
dominated elliptical galaxy, a consequence of a low global baryon
fraction or a high concentration of dark, relative to luminous,
matter. Although one must therefore exercise caution in generalizing
from our results, the most stringent constraints on alternative models
for dark halo structure, that presumably universally apply, emerge.
Results, in progress, of similar investigations for other galaxies
(e.g., NGC 1399, NGC 4472) are anticipated with great interest.

\acknowledgments

We are extremely grateful to S. Khan, J. R. Peterson, and the {\it
XMM-Newton} RGS team for providing {\it XMM-Newton} results for NGC
4636 prior to publication, and greatly appreciate the assistance of
Lorella Angelini and Keith Arnaud with image and spectral analysis of
{\it Chandra} data.

\clearpage


\clearpage

\begin{deluxetable}{ccccccc}
\tablecaption{Results of {\it Chandra} Spectral Fitting$\tablenotemark{a}$}
\tablewidth{0pt}
\tablehead{
\colhead{annulus} & 
\colhead{$kT$} & 
\colhead{O} & 
\colhead{Fe}  &
\colhead{Si} & 
\colhead{$N_H$} & 
\colhead{${{\chi}_{\nu}}^2$} \\
}
\startdata  
$0-10''$  & 0.509 & 0.078 & 0.413 & 0.664 & 0.048 & 299/189 \\ 
$10-20''$ & 0.578 & 0.173 & 0.64  & 0.76  & 0.033 & 229/178 \\ 
$20-40''$ & 0.616 & 0.113 & 0.776 & 0.862 & 0.033 & 308/178 \\ 
$40-80''$ & 0.68 & 0.02 & 0.84 & 1.127 & 0.056 &
359/171\tablenotemark{b} \\ 
\enddata

\tablenotetext{a}{Temperature, $kT$, in keV; abundances relative to
solar; column density, $N_H$ in $10^{22}$ cm$^{-2}$; ${\chi_{\nu}}^2$
is $\chi^2$ per degree of freedom}
\tablenotetext{b}{residuals at 1.022 keV}
\end{deluxetable}

\clearpage

\begin{deluxetable}{cccccccc}
\tablecaption{Mass Model Best-fit Parameters}
\tablewidth{0pt}
\tablehead{
\colhead{$\zeta_1$} & 
\colhead{$r_{\rm dark,1}$\tablenotemark{a}} & 
\colhead{${{M_{\rm dark,1}}\over {M_{\rm stars}}}$\tablenotemark{b}}  & 
\colhead{$\zeta_2$} & 
\colhead{$r_{\rm dark,2}$\tablenotemark{a}} & 
\colhead{${{M_{\rm dark,2}}\over {M_{\rm stars}}}$\tablenotemark{b}}  &
\colhead{${{M_{\rm stars}}\over {L_V}}$} & 
\colhead{$\langle\Delta kT\rangle$\tablenotemark{c}} \\
}
\startdata  
1.0 &  32 &  7.6 & ... &  ... &  ... & 5.5 & 10.6 \\ 
1.5 &  55 & 13.7 & ... &  ... &  ... & 3.2 &  9.2 \\ 
2.0 & 780 & 18.8 & ... &  ... &  ... & 1.5 &  201 \\ 
0.0 &  12 &  6.2 & ... &  ... &  ... & 6.6 & 17.2 \\ 
0.0\tablenotemark{d} 
    & 4.5 &  6.1 & ... &  ... &  ... & 6.3 & 13.0 \\ 
1.0 &  68 &  4.8 & ... &  ... &  ... & 8.0\tablenotemark{e} & 48.6 \\ 
0.0 &  17 &  5.1 & ... &  ... &  ... & 8.0\tablenotemark{e} & 32.9 \\ 
0.0\tablenotemark{d}
    & 7.5 &  4.7 & ... &  ... &  ... & 8.0\tablenotemark{e} & 32.2 \\ 
1.0 &  41 & 10.3 & 0.0 & 0.99 & 0.83 & 3.9 &  6.1 \\ 
1.0 &  54 &  6.3 & 0.0 &  3.6 &  1.4 & 5.4\tablenotemark{e} &  8.3 \\ 
1.5 &  74 & 12.8 & 0.0 &  2.2 & 0.77 & 3.2 &  8.2 \\ 
2.0 & 6.6 & 0.01 & 0.0 &   12 &  6.2 & 6.6 & 17.4 \\
1.0 &  44 & 10.3 & 0.0\tablenotemark{d}
                       & 0.44 &  6.2 & 2.7 &  6.2 \\ 
1.5 &  67 & 12.4 & 0.0\tablenotemark{d}
                       & 0.85 &  2.3 & 3.0 &  7.5 \\ 
2.0 &  29 & 0.01 & 0.0\tablenotemark{d}
                       &  4.7 &  6.0 & 6.4 & 12.9 \\  
\enddata

\tablenotetext{a}{in kpc}
\tablenotetext{b}{evaluated at $200r_{\rm break}=10.7'=46.7$ kpc}
\tablenotetext{c}{rms deviation in eV}
\tablenotetext{d}{pseudo-isothermal model}
\tablenotetext{e}{fixed}

\end{deluxetable}

\clearpage

\begin{deluxetable}{cccccccc}
\tablecaption{Best-fit Characteristics for Selected
Models\tablenotemark{a}}
\tablewidth{0pt}
\tablehead{
\colhead{$\zeta_1$} & 
\colhead{$\zeta_2$} & 
\colhead{$\rho(r_{\rm break})$} & 
\colhead{$f_{\rm dark}(r_{\rm break})$} & 
\colhead{$\rho(r_{\rm e})$} & 
\colhead{$f_{\rm dark}(r_{\rm e})$} & 
\colhead{${{M_{\rm total}}\over {L_V}}(200r_{\rm break})$} &
\colhead{$\Delta_{V/2}$} \\
}
\startdata  
1.0 & ... &  34 & 0.091 & 0.11 & 0.60 & 47 & $6.9\times 10^5$ \\
1.5 & ... &  36 & 0.49  & 0.12 & 0.77 & 47 & $3.2\times 10^6$ \\
0.0 & ... &  37 & 0.005 & 0.11 & 0.51 & 47 & $4.2\times 10^5$ \\
0.0\tablenotemark{b}
    & ... &  35 & 0.005 & 0.12 & 0.56 & 45 & $5.5\times 10^5$ \\
1.0\tablenotemark{c}
    & ... &  45 & 0.037 & 0.11 & 0.40 & 47 & $1.7\times 10^5$ \\
0.0\tablenotemark{c}
    & ... &  44 & 0.002 & 0.11 & 0.38 & 49 & $2.2\times 10^5$ \\
0.0\tablenotemark{b,c}
    & ... &  44 & 0.002 & 0.11 & 0.39 & 46 & $2.2\times 10^5$ \\
1.0 & 0.0 &  27 & 0.20  & 0.12 & 0.72 & 43 & $2.5\times 10^6$ \\
1.0\tablenotemark{d}
    & 0.0 &  32 & 0.064 & 0.11 & 0.62 & 41 & $8.3\times 10^5$ \\
1.5 & 0.0 &  34 & 0.46  & 0.12 & 0.78 & 44 & $3.2\times 10^6$ \\
1.0 & 0.0\tablenotemark{b}
          &  23 & 0.32  & 0.12 & 0.81 & 44 & $6.3\times 10^6$ \\
1.5 & 0.0\tablenotemark{b}
          &  32 & 0.46  & 0.12 & 0.79 & 44 & $4.0\times 10^6$ \\
\enddata

\tablenotetext{a}{total mass densities (in M$_{\odot}$ pc$^{-3}$) and
dark matter fractions evaluated at $r_{\rm break}=233$ pc and
effective radius $r_e=7.8$ kpc}
\tablenotetext{b}{pseudo-isothermal model} 
\tablenotetext{c}{${{M_{\rm stars}}\over {L_V}}$ fixed at 8} 
\tablenotetext{d}{${{M_{\rm stars}}\over {L_V}}$ fixed at 5.4}

\end{deluxetable}

\clearpage


\begin{figure}
\plotone{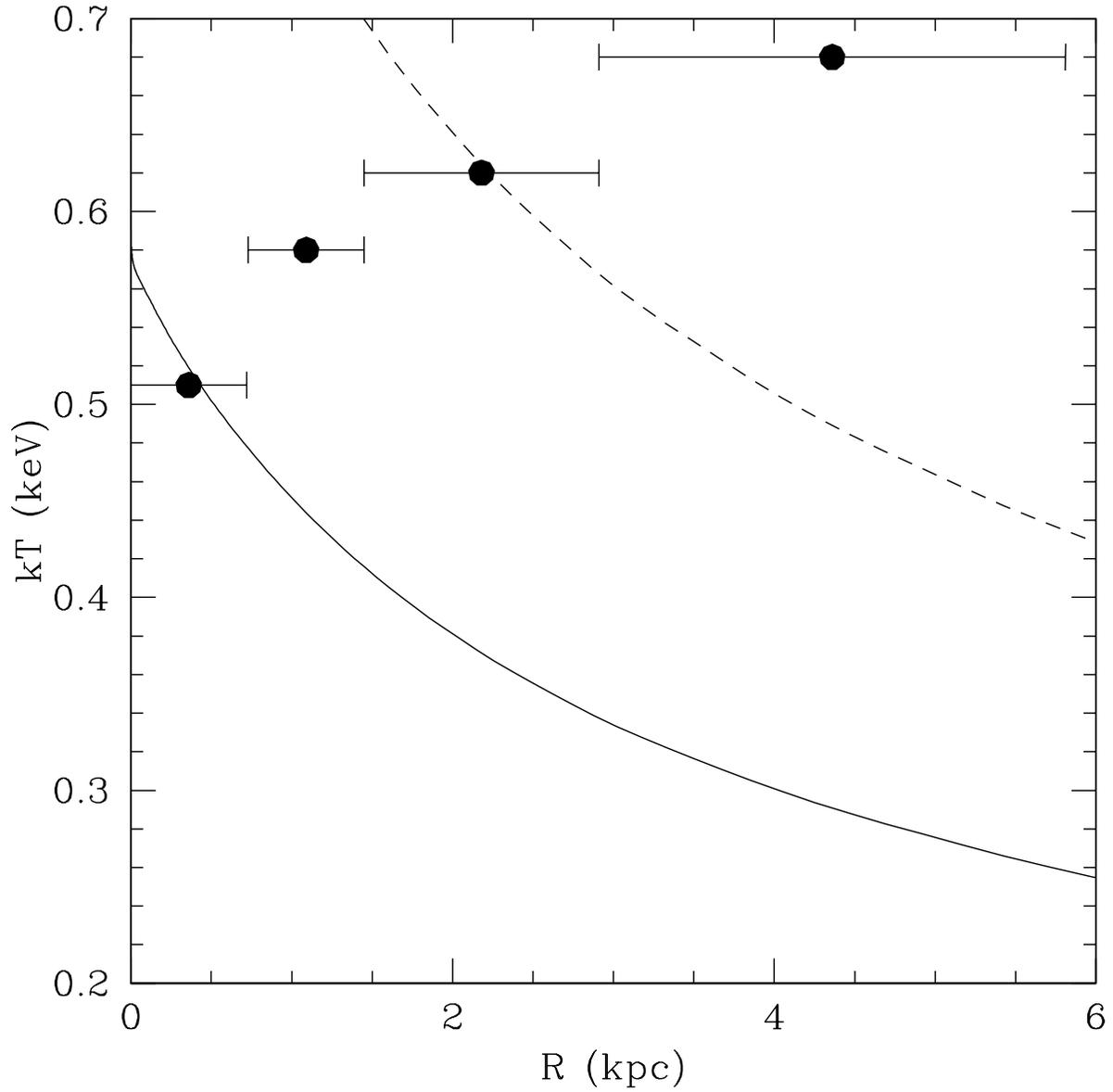}
\figcaption{{\it Chandra}, $<80''$ temperature profile (filled
circles, with errorbars representing the extraction annuli), and
predictions for models with constant mass-to-light ratios, $M/L_V=8$
(solid curve) and $M/L_V=13.5$ (dashed curve).}
\end{figure}

\begin{figure}
\plotone{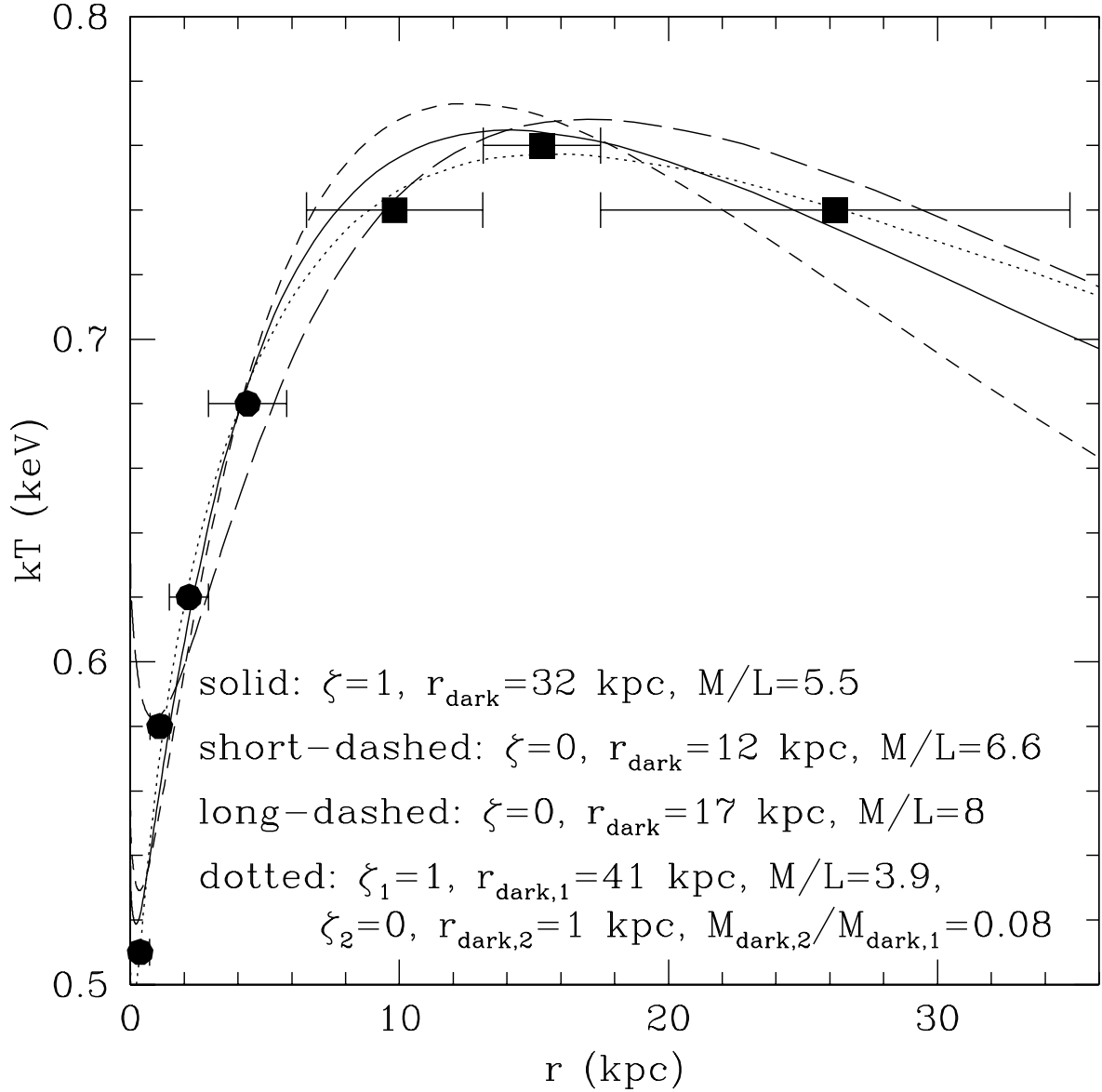}
\figcaption{{\it Chandra} (filled circles) and {\it XMM-Newton}
(filled squares) temperatures and predictions for the following
best-fitting models: one-component $\zeta=1$ model (solid curve),
one-component $\zeta=0$ model (short-dashed curve), one-component
$\zeta=0$ model with $M_{\rm stars}/L_V$ fixed at 8 (long-dashed
curve), two-component model with $(\zeta_1,\zeta_2)=(1,0)$ (dotted
curve).}
\end{figure}

\begin{figure}
\plotone{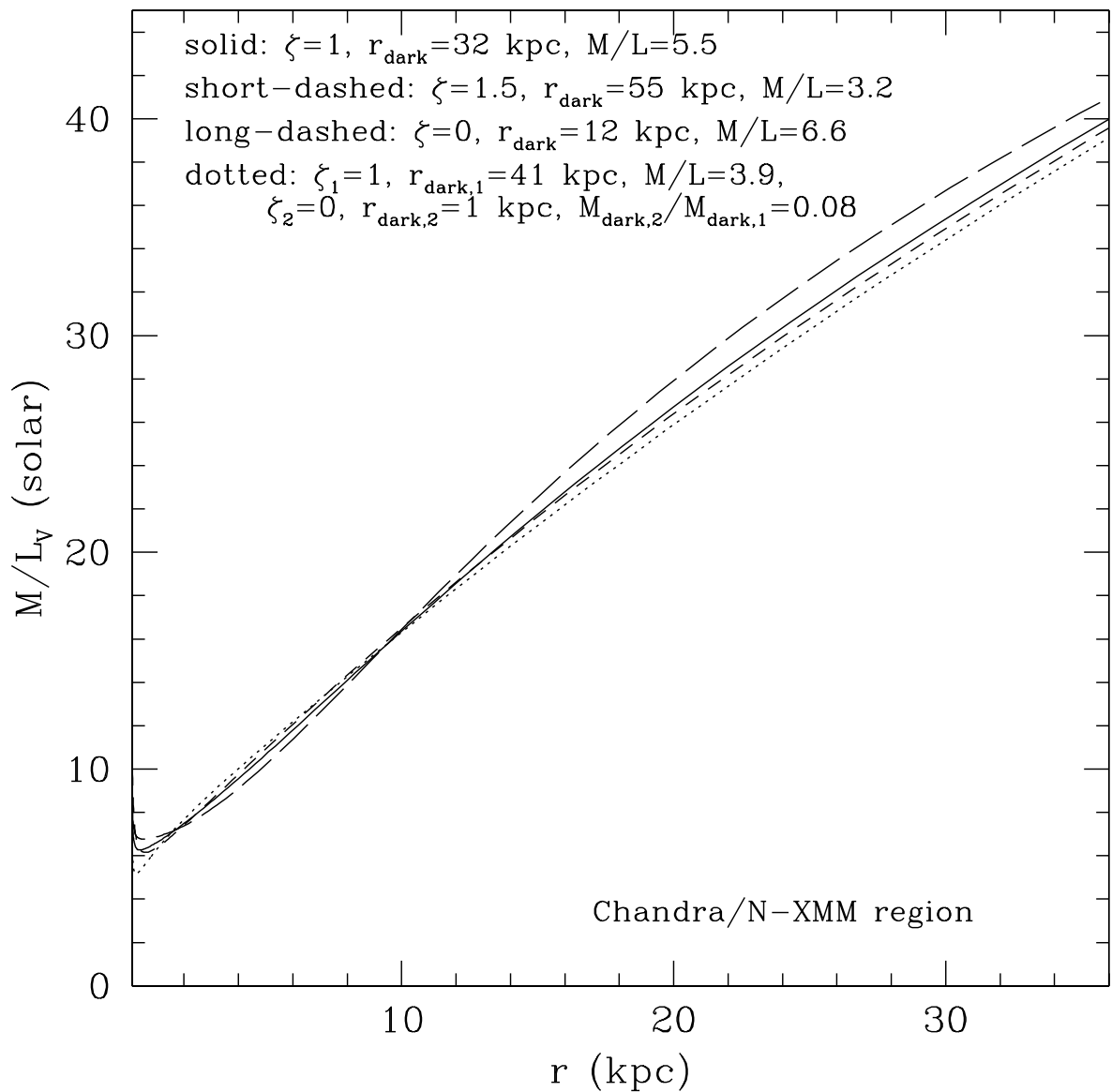}
\figcaption{Total mass-to-light ratios for the following best-fitting
models: one-component $\zeta=1$ model (solid curve), one-component
$\zeta=1.5$ model (short-dashed curve), one-component $\zeta=0$ model
(long-dashed curve), two-component model with
$(\zeta_1,\zeta_2)=(1,0)$ (dotted curve).}
\end{figure}

\begin{figure}
\plottwo{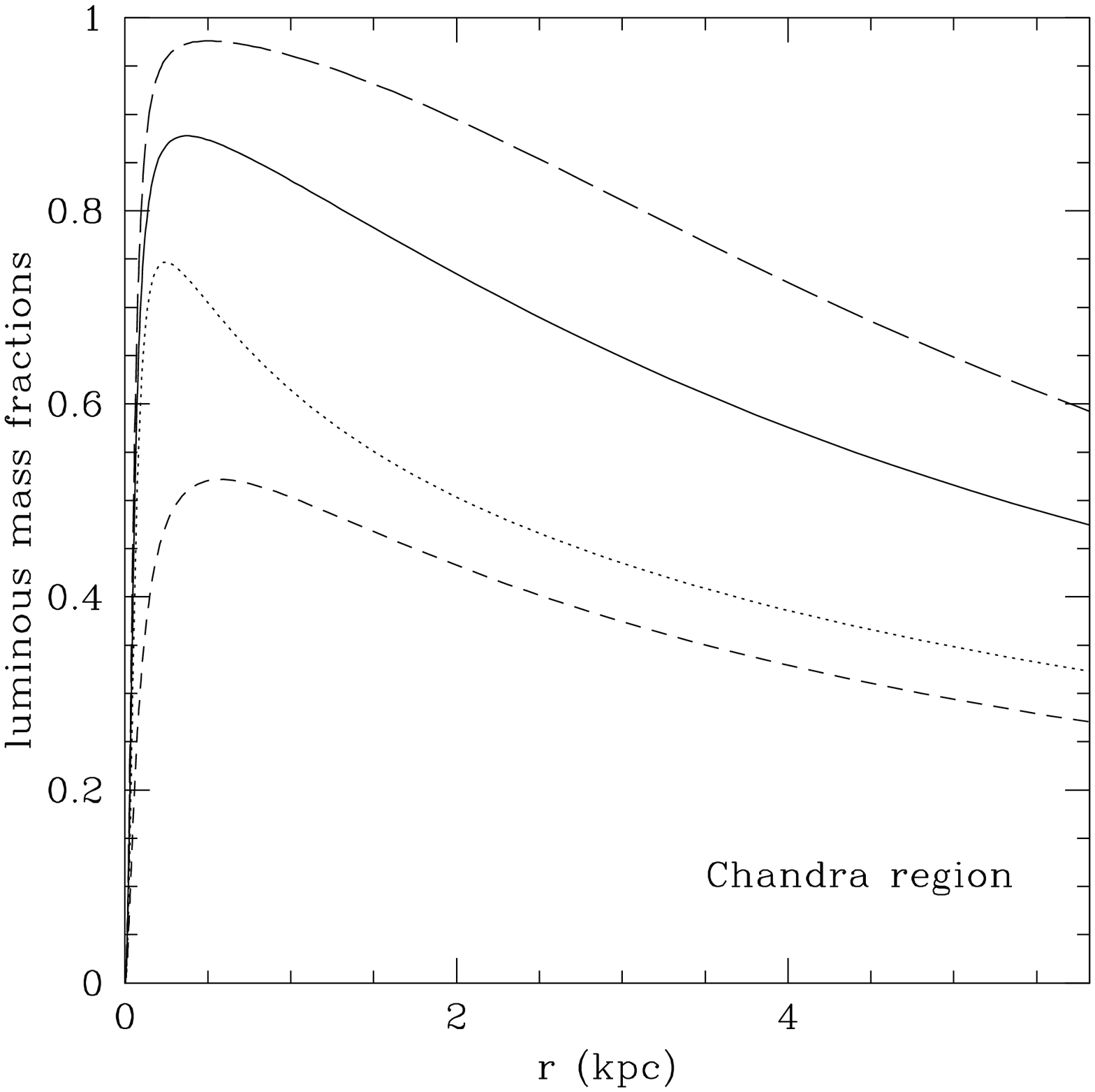}{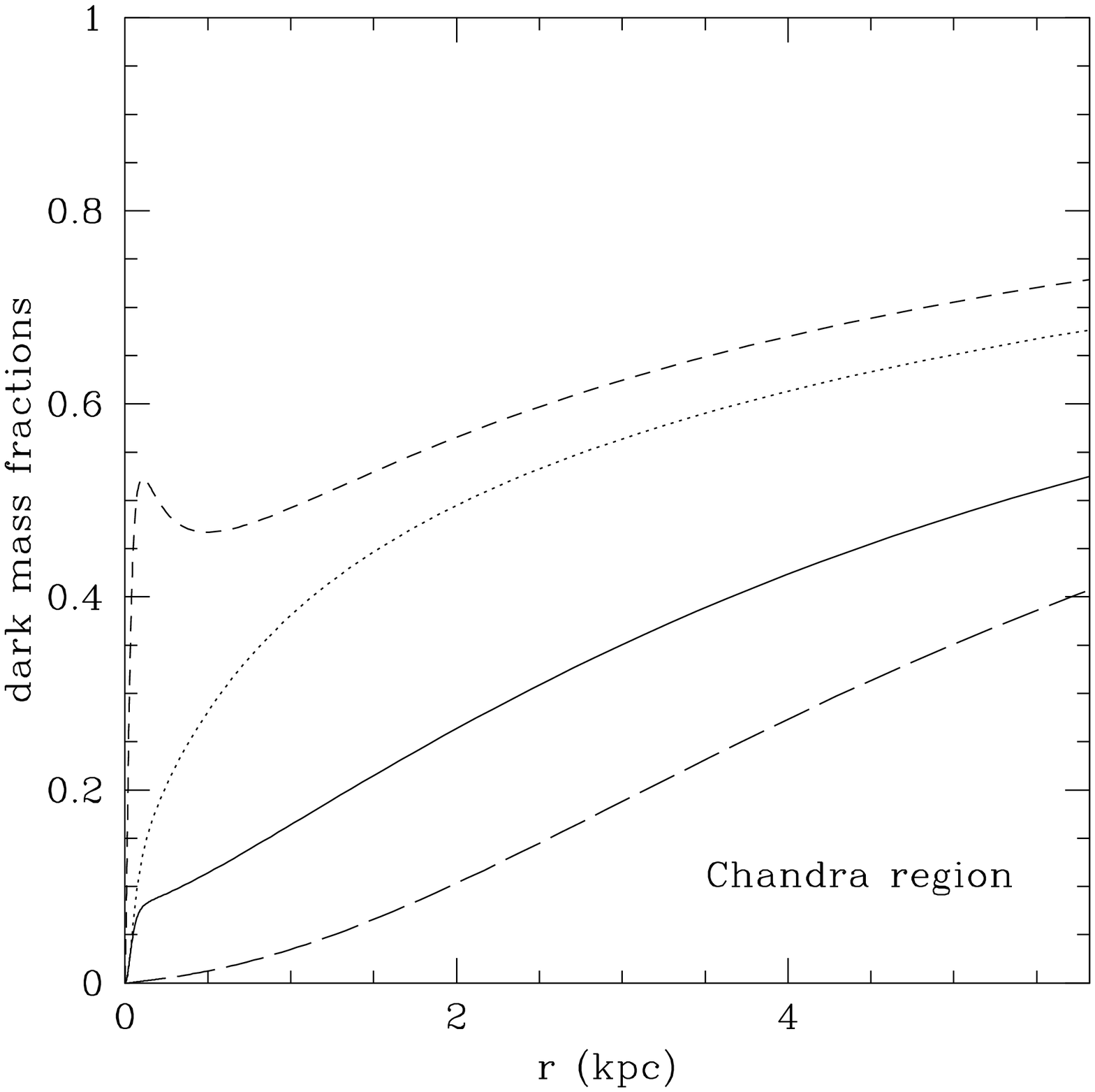}
\plotone{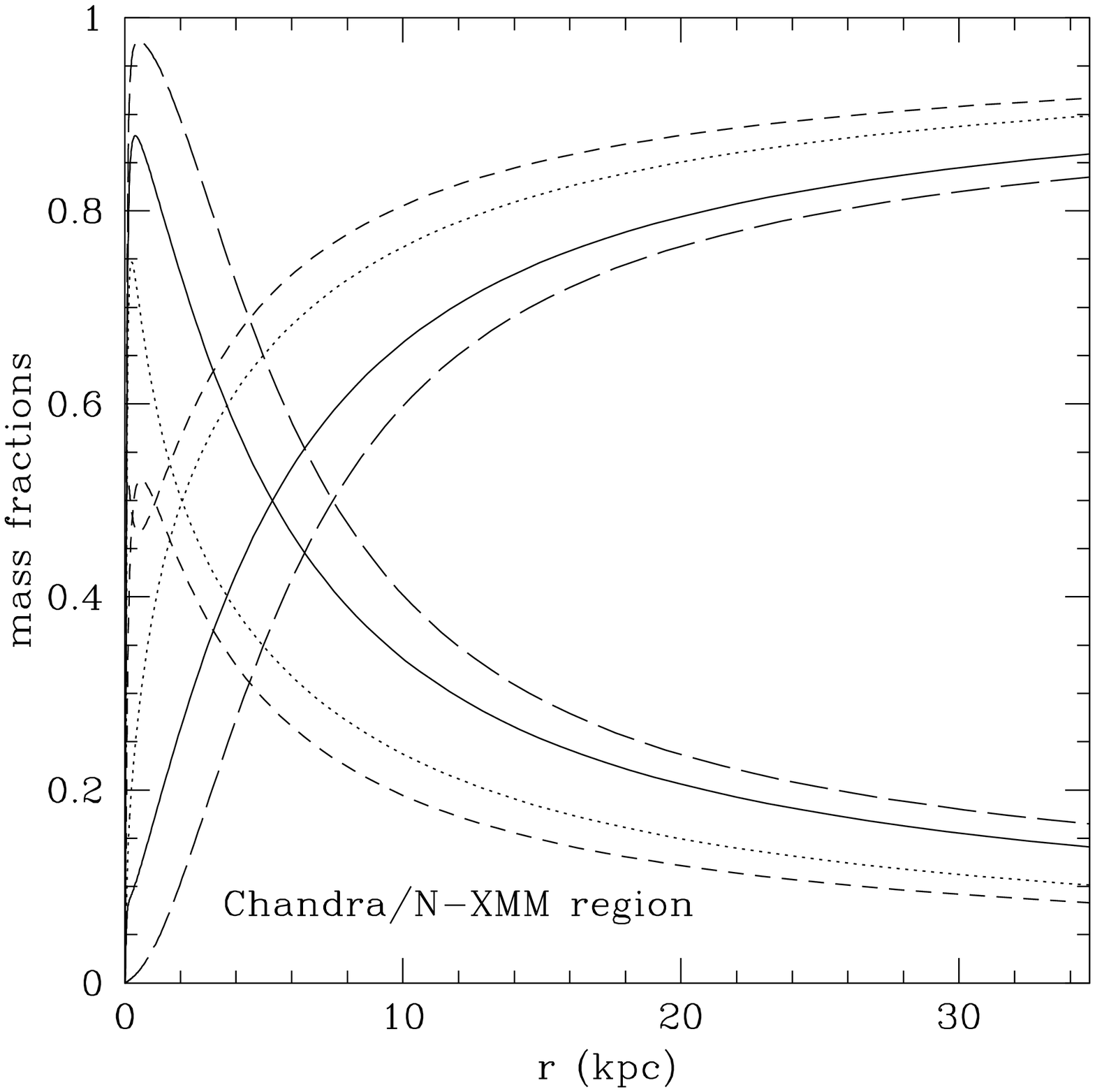}
\figcaption{Same as Figure 3 for the luminous (a) and dark (b)
matter fractions on the {\it Chandra} scale, and on the 
{\it XMM-Newton} scale (c).}
\end{figure}

\begin{figure}
\plottwo{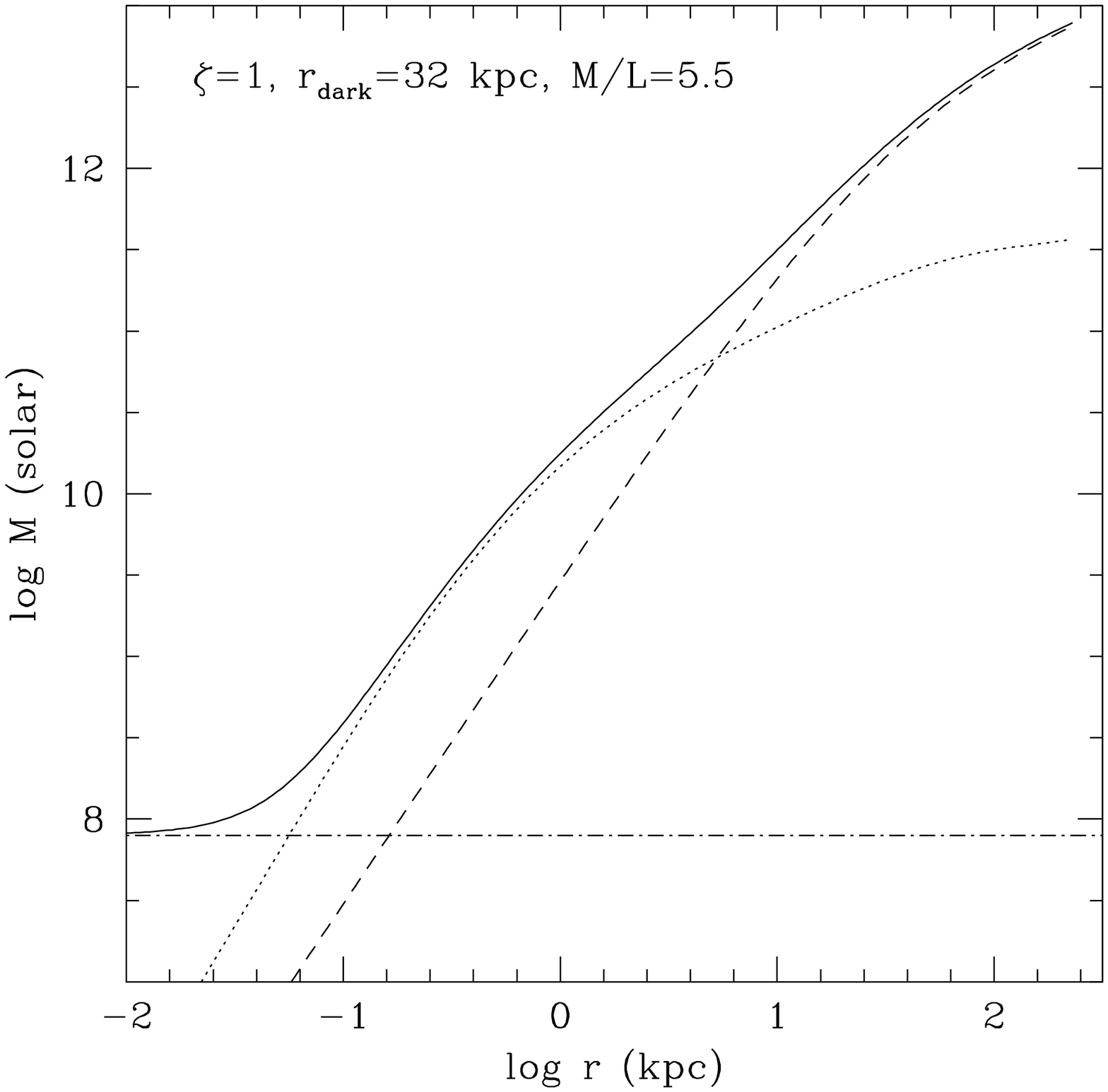}{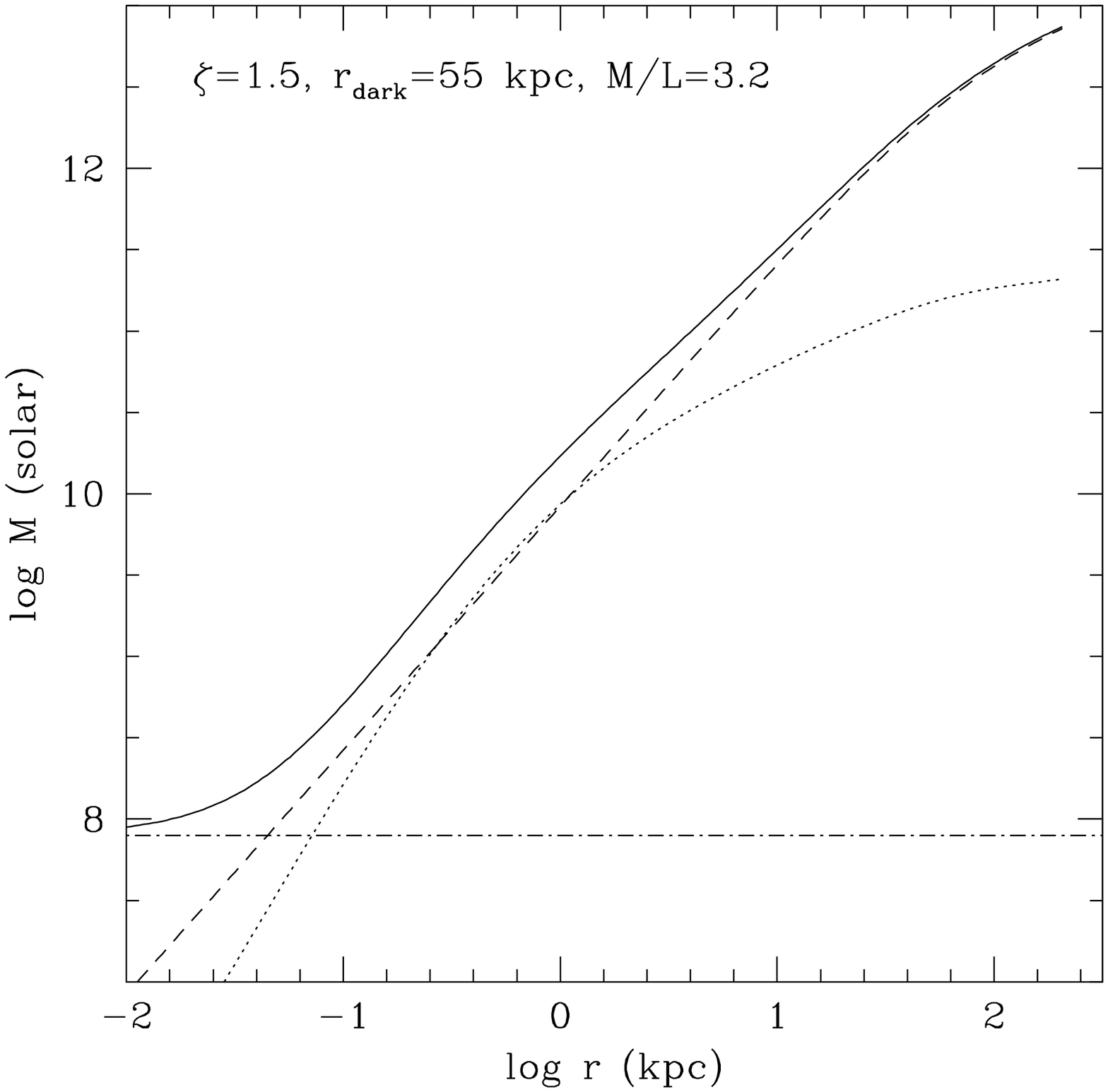}
\plotone{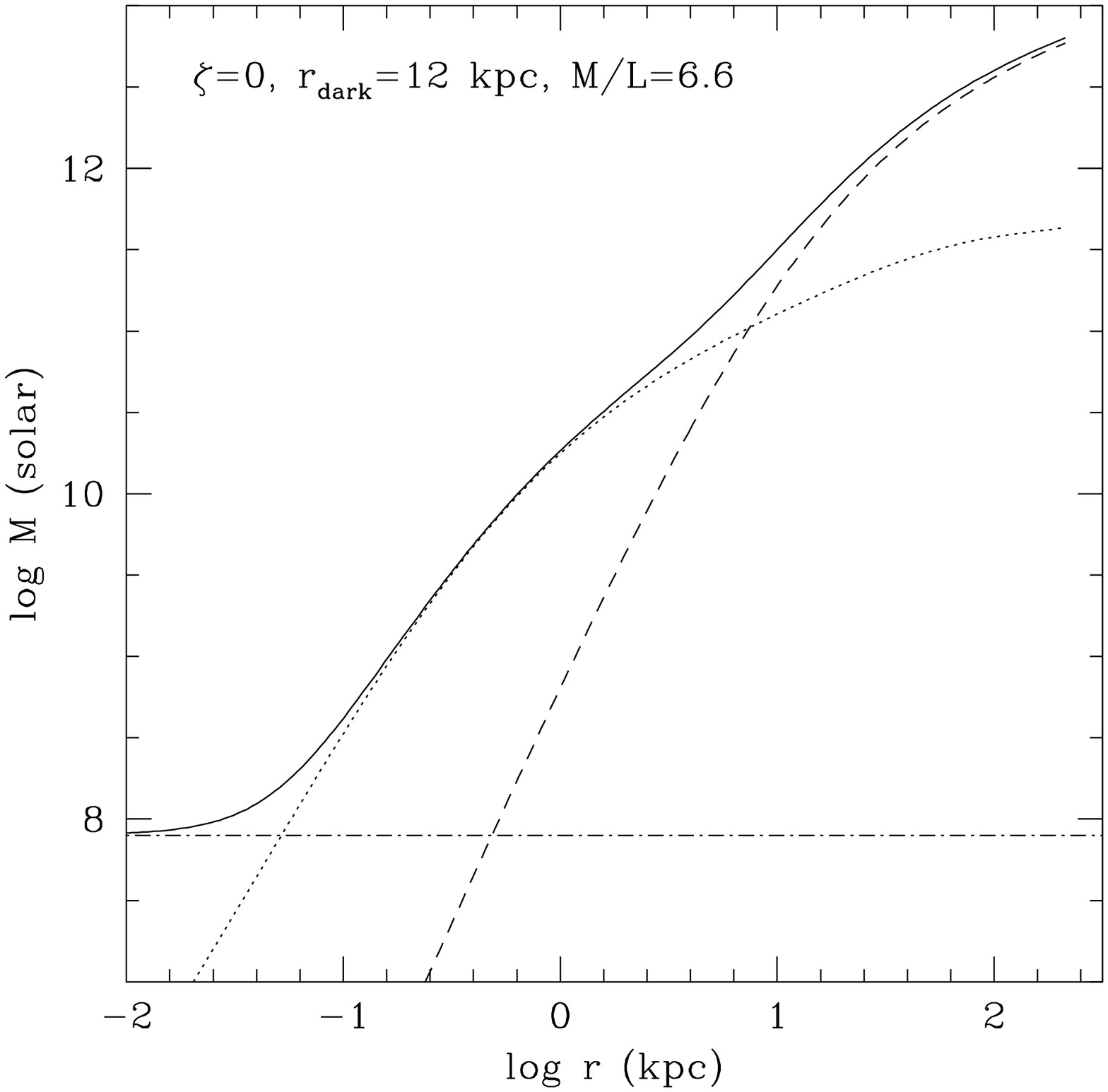}
\figcaption{Mass decomposition, showing stellar (dotted curve), dark
matter (dashed curve), SMBH (dot-dashed curve), and total (solid)
contributions for the best-fitting one-component $\zeta=1$ model (a),
the best-fitting one-component $\zeta=1.5$ model (b), and the
best-fitting one-component $\zeta=0$ model (c).}
\end{figure}

\begin{figure}
\plottwo{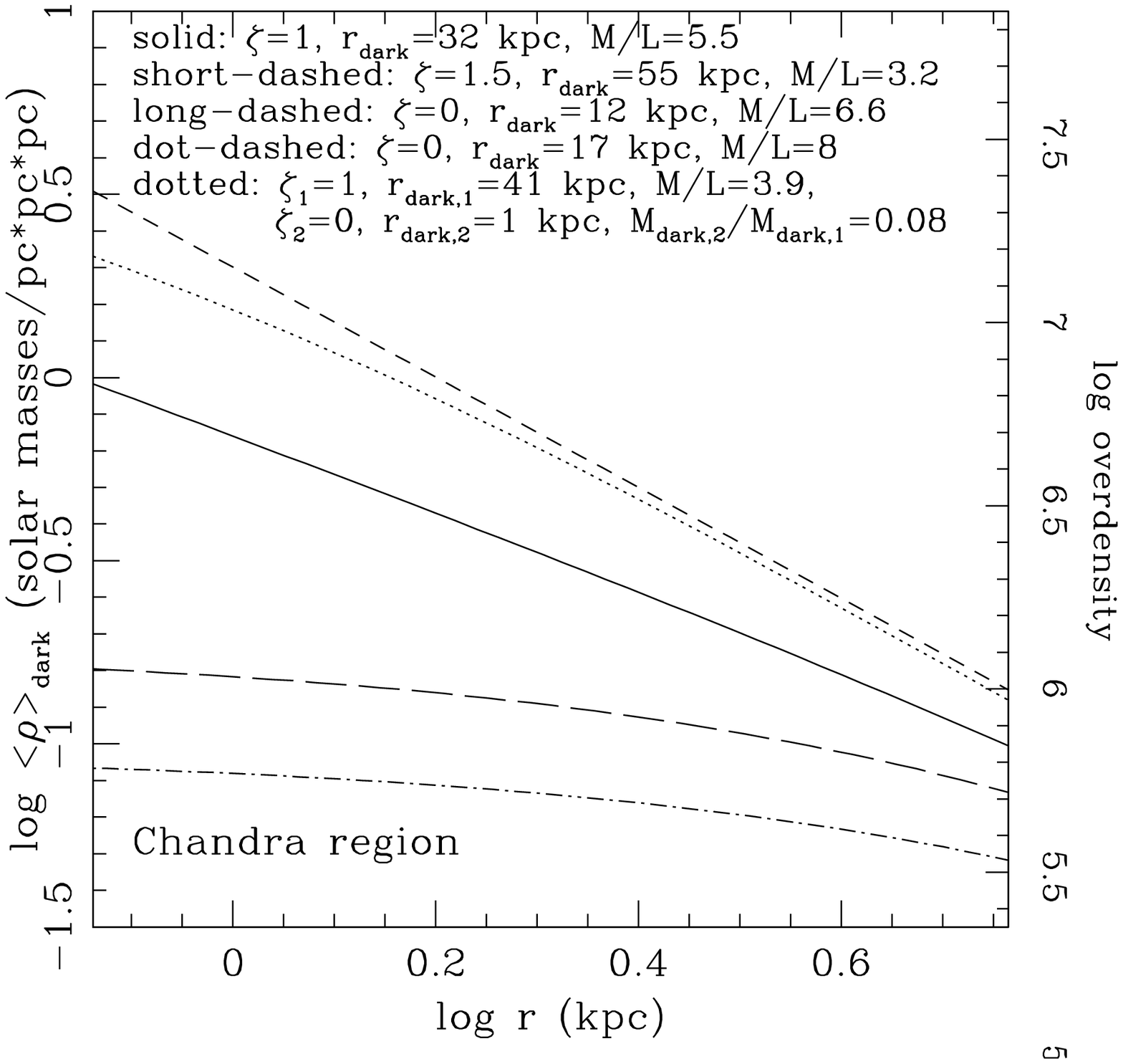}{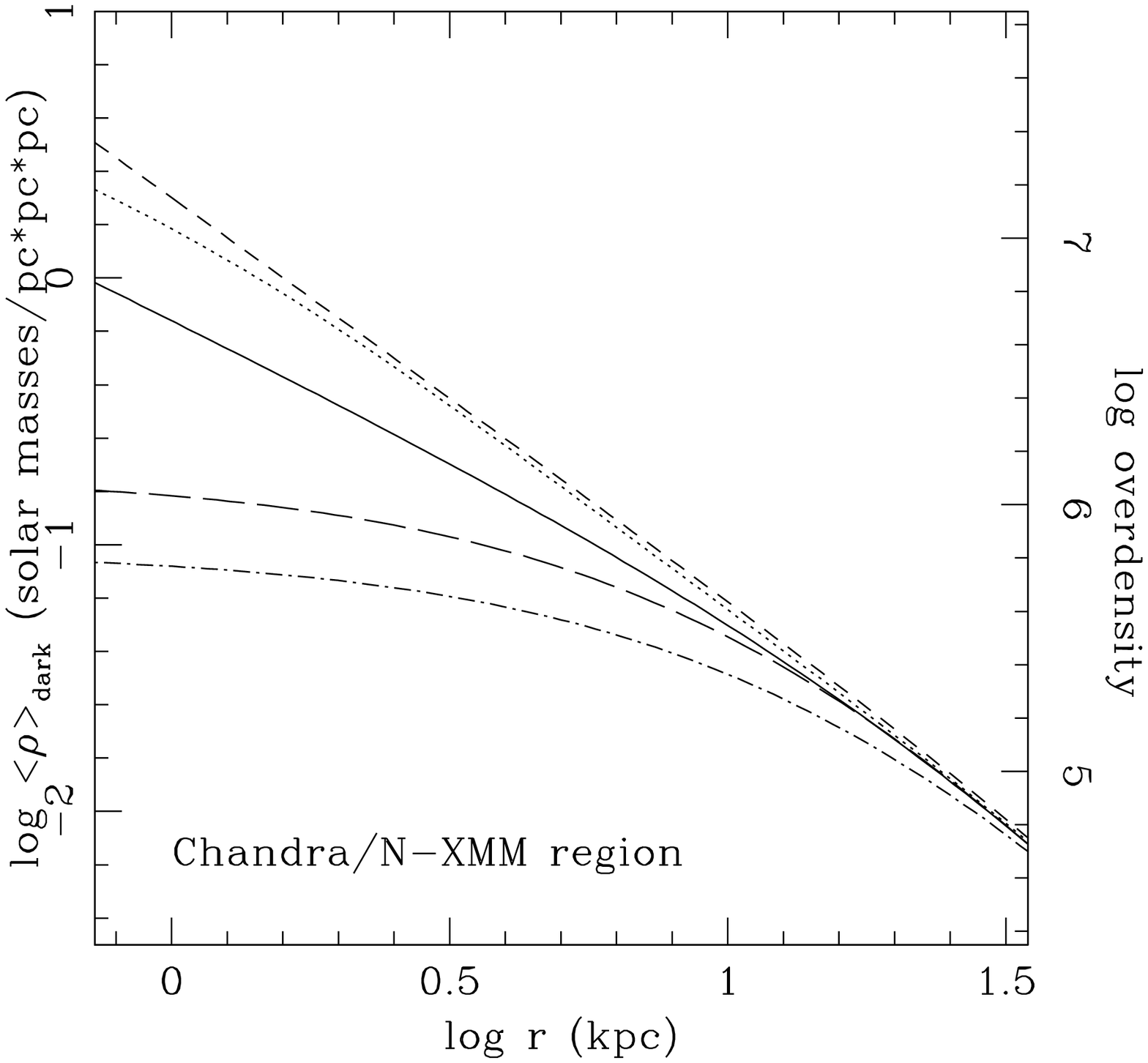}
\figcaption{Average enclosed dark matter density distribution in the
{\it Chandra} (a) and {\it XMM-Newton} (b) regions. Curves are as in
Figures 3 and 4, with an additional (dot-dashed) curve for the
best-fit one-component $\zeta=0$ model with $M_{\rm stars}/L_V$ fixed
at 8. Corresponding scale in units of overdensity is also shown.}
\end{figure}

\begin{figure}
\plottwo{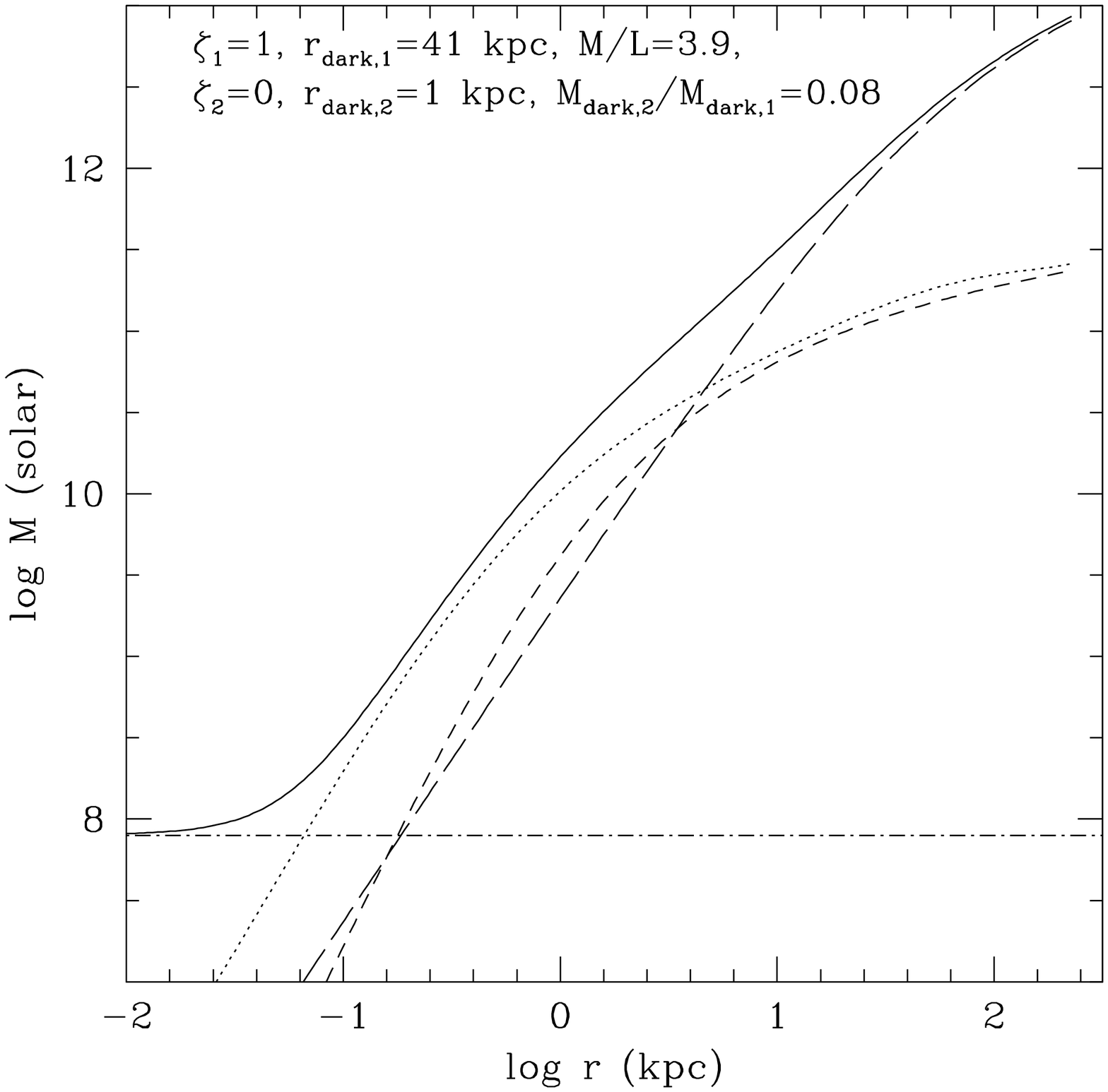}{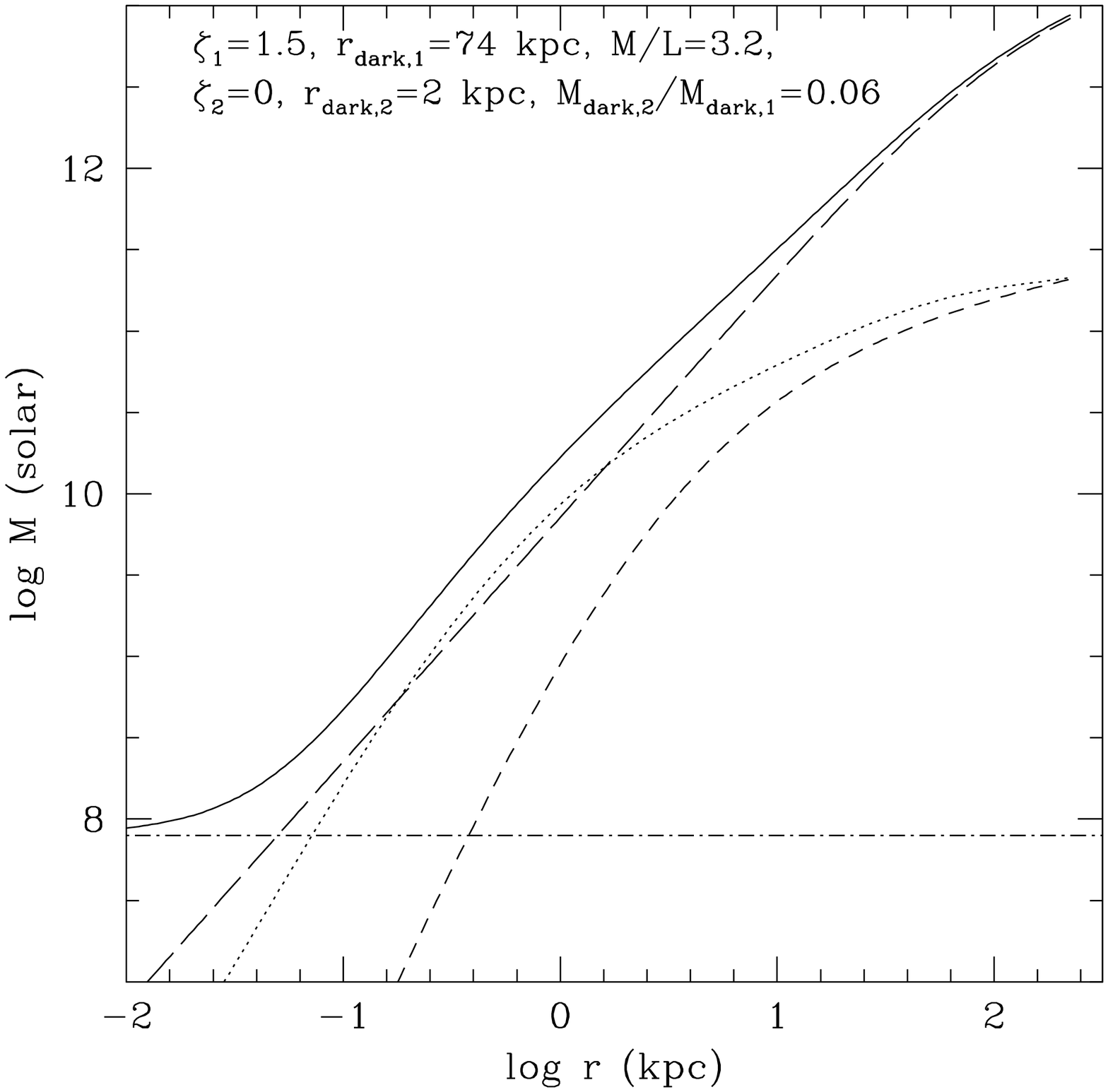}
\figcaption{Mass decomposition as in Figure 5 for the best-fit
two-component models with $(\zeta_1,\zeta_2)=(1,0)$ (a) and
$(\zeta_1,\zeta_2)=(1.5,0)$ (b). The short-dashed curves show the flat
dark matter core component, the long-dashed curves the cuspy core
component.}
\end{figure}

\begin{figure}
\plotone{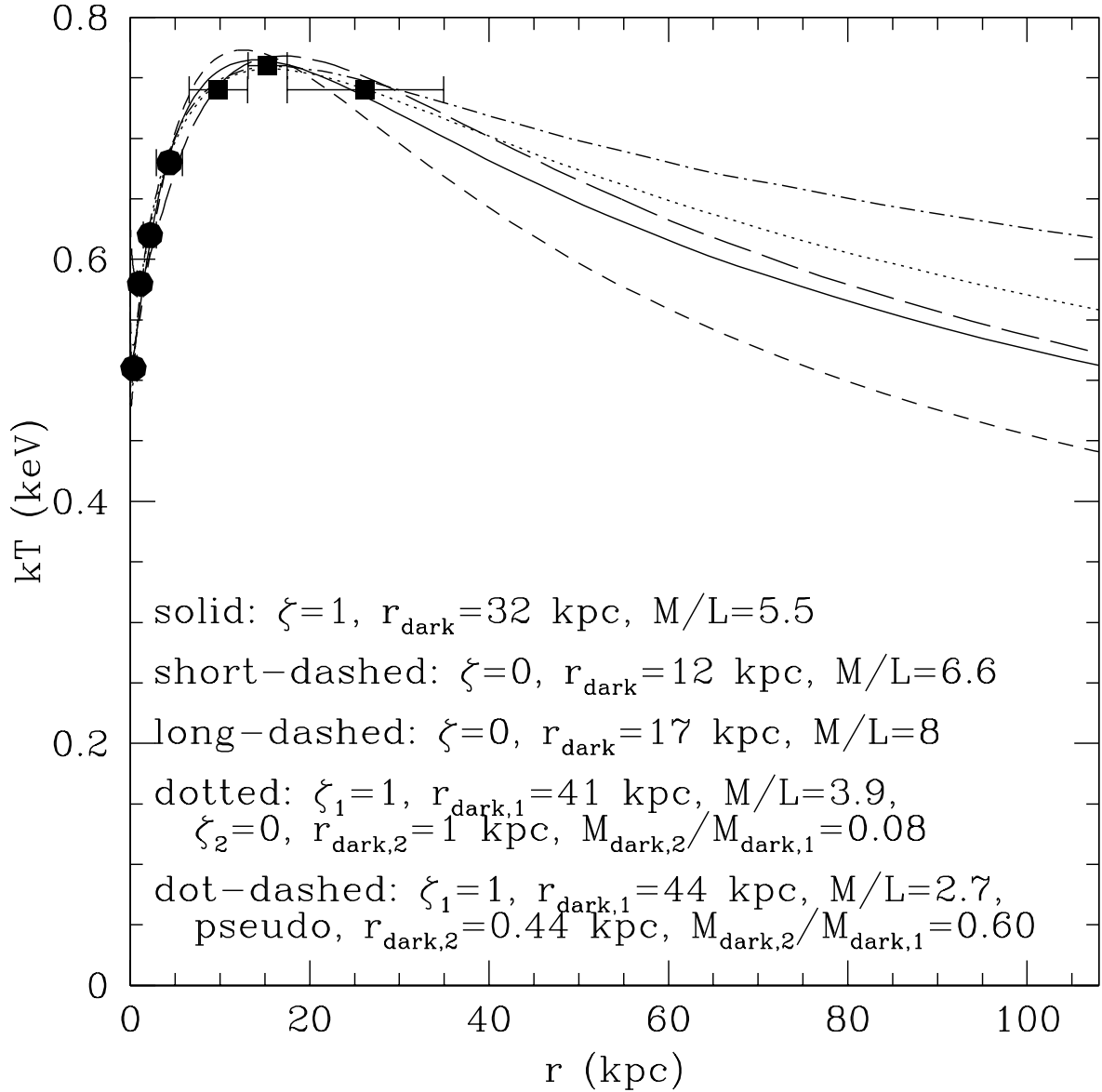}
\figcaption{Measured and predicted temperature profiles as in Figure
2, extended to $>100$ kpc, and with an extra (dot-dashed) curve for
best-fit composite model with NFW ($\zeta=1$) and pseudo-isothermal
dark matter components.}
\end{figure}

\begin{figure}
\plotone{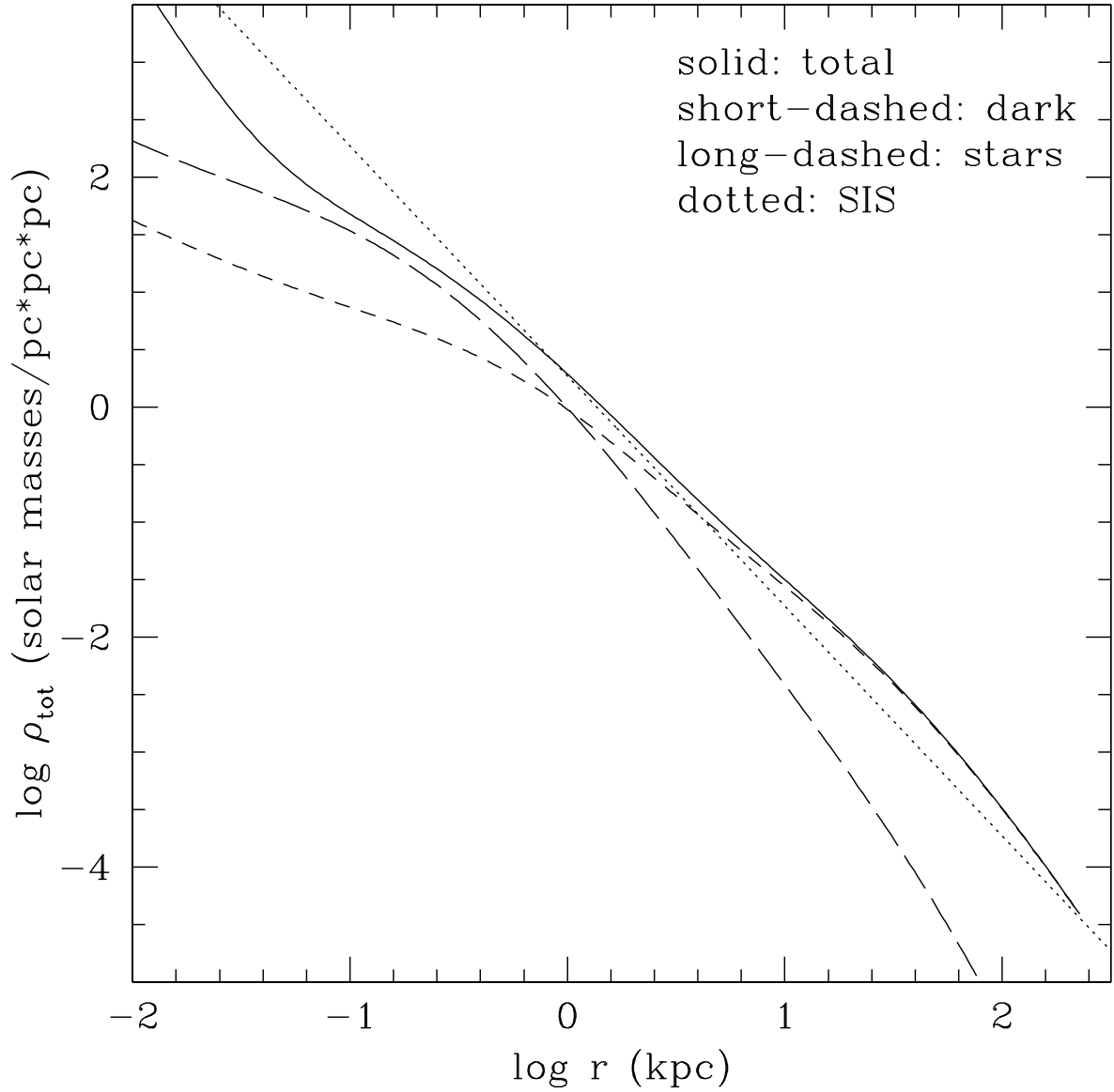}
\figcaption{Comparison of mass density distribution (stellar:
long-dashed curve, dark matter: short-dashed curve, total:solid curve)
for best-fit two-component model with $(\zeta_1,\zeta_2)=(1,0)$ and
singular isothermal sphere of identical virial mass (dotted curve).}
\end{figure}

\end{document}